\documentclass[12pt]{article}
\usepackage{ctex}
\usepackage{amssymb,amsmath,color,mathdots}
\usepackage[colorlinks,linkcolor=blue]{hyperref}
\usepackage{geometry}
\usepackage{tabularx}
\usepackage{bm}
\usepackage{authblk}
\usepackage[square, comma, sort&compress, numbers]{natbib}

\title{\bf Constructions for several  classes of few-weight linear codes and their applications}	
\author{\small Canze Zhu}
\author{\small Qunying Liao
	\thanks{Corresponding author.
		
		{~E-mail. qunyingliao@sicnu.edu.cn (Q. Liao), ~canzezhu@163.com (C. Zhu).}	
		
		{~Supported by National Natural Science Foundation of China (Grant No. 12071321).}}
}
\affil[]{\small(College of Mathematical Science, Sichuan Normal University, Chengdu Sichuan, 610066)}
\date{}
\newtheorem{corollary}{Corollary}[section]
	\newtheorem{theorem}{Theorem}[section]
	
	\newtheorem{lemma}{Lemma}[section]
	\newtheorem{example}{Example}[section]

	\newtheorem{remark}{Remark}[section]

	\catcode`@=11 \@addtoreset{equation}{section} \catcode`@=12
\begin{document}
	\maketitle
	{\bf Abstract.}
	{\small
		In this paper, for any odd prime $p$ and  an integer $m\ge 3$, several classes of linear codes with $t$-weight $(t=3,5,7)$ are obtained based on some defining sets, and then their complete weight enumerators   are determined explicitly by employing Gauss sums and quadratic character sums. Especially for $m = 3$, a class of  MDS codes with parameters $[p,3,p-2]$ are obtained. Furthermore, some of these codes can be suitable for applications in secret sharing schemes and $s$-sum sets  for any odd $s>1$. }\\

		

	\section{Introduction}Let $\mathbb{F}_{p^m}$ be the finite field with $p^m$ elements and $\mathbb{F}_{p^m}^*=\mathbb{F}_{p^m}\backslash \{0\}$, where $p$ is an odd prime and $m$ is a positive integer. An $[n,k,d]$ linear code $\mathcal{C}$ over $\mathbb{F}_p$ is a $k$-dimensional subspace of $\mathbb{F}_p^n$ with minimum (Hamming) distance $d$ and length $n$. 
	The dual code of $\mathcal{C}$ is defined as 
	$$	 \mathcal{C}^{\perp}=\{\mathbf{c}^{'}\in\mathbb{F}_p^n~|~\langle\mathbf{c}^{'},\mathbf{c}\rangle=0~ \text{for any}~\mathbf{c}\in\mathcal{C}\}.$$
	Clearly, the dimension of $\mathcal{C}^{\perp}$ is $n-k$. A linear code $\mathcal{C}$ is said to be projective if the minimum distance of $\mathcal{C}^{\perp}$  is not less than $3$.
	
	Let $A_i\ (i=0,1,\ldots,n)$ be the number of codewords with Hamming weight $i$ in $\mathcal{C}$, then the weight enumerator  of $\mathcal{C}$ is defined by the polynomial
	$$1+A_1z+\cdots+A_nz^n. $$ $\mathcal{C}$ is called a $t$-weight code if the number of nonzero $A_i$ in the sequence $(A_1,\ldots,A_n)$ is equal to $t$. In addition, the complete weight enumerator for a codeword $\mathbf{c}$ is the monomial
	$$w(\mathbf{c})=w_0^{t_0}w_1^{t_1}\cdots w_{p-1}^{t_{p-1}}$$
	in the variables $w_0,w_1,\ldots,w_{p-1}$, where $t_i\ (0\le i\le p-1)$ denotes the number of components of $\mathbf{c}$ equal to $i$. The complete weight enumerator for $\mathcal{C}$ is defined to be
	\begin{center}
		$W(\mathcal{C})=\sum\limits_{\mathbf{c}\in\mathcal{C}}w(\mathbf{c})$.
	\end{center} 
	
	The complete weight enumerator is an important parameter for a linear code, obviously, the weight distribution  can be deduced from the complete weight enumerator. In addition, the weight distribution for a code contains some important information on its error-correcting capability and the error probability of its error detection and correction with respect to some algorithms \cite{KT2007}. It is well-known that few-weight linear codes have better applications in secret sharing schemes \cite{JY2006,CC2005},  association schemes \cite{AC1984}, authentication codes \cite{CD2005}, $s$-sum sets \cite{SS,SK}, and so on. Several constructions for few-weight codes, mainly based on special defining sets, several few-weight linear codes have been obtained \cite{KD2014, SY2015, ZH2015, ZH2016, ZH20161, ZH20162, GJ2019, CL2016, GL2018, CT2016, ZZ2015,Z,YK, CS2019}. 
	
	In 2015, Ding et al. gave a generic construction for few-weight linear codes \cite{KD2015}. Let $\mathrm{Tr}$ be the absolute trace function from $\mathbb{F}_{p^{m}}$ to $\mathbb{F}_p$ and 	$D=\{ d_1,\ldots, d_n \}\subseteq\mathbb{F}_{p^{m}}^*$, the $p$-ary linear code is defined by
	\begin{align*}
	\mathcal{C}_{D}=\big\{\mathbf{c}(b)=\big(\mathrm{Tr}(bx)\big)_{x\in D}~\big{|}~ b\in\mathbb{F}_{p^{m}}\big\}.
	\end{align*}
	
	In 2017, Yang et al. proposed a similar construction \cite{SY2015}, they defined the linear code
	\begin{align} \label{C1}
	\mathcal{C}_{D}=\big\{\mathbf{c}(b)=\big(\mathrm{Tr}(bx^2)\big)_{x\in D}~\big{|}~ b\in\mathbb{F}_{p^{m}}\big\}
	\end{align}
	 with
	 \begin{align*}
	 	D=\{x\in\mathbb{F}_{p^m}^{*}~|~\mathrm{Tr}(x)=0\},
	 \end{align*}
 then a class of three-weight codes are obtained and the corresponding complete weight enumerators are determined. After that, Wang et al.  obtained several classes of three-weight linear codes  from $(\ref{C1})$ by choosing the defining set 
\begin{align*}
D=\{x\in\mathbb{F}_{p^m}^{*}~|~\mathrm{Tr}(x)=\lambda\},
\end{align*}
where $\lambda\in\mathbb{F}_{p}^*$ \cite{Wq}.

	In this paper, 
	for any fixed $a\in\mathbb{F}_{p^m}\backslash\mathbb{F}_p$,	we  investigate the code	$\mathcal{C}_{D_a}$ in $(\ref{C1})$ with the defining set 
	\begin{align*}
	D_{a}=\{ x\in\mathbb{F}_{p^m} ~\big|~\mathrm{Tr}(x)=1~\text{and}~\mathrm{Tr}(ax)=0\}.
	\end{align*}
	
		By taking some special $a$, we can obtain several classes of linear codes with $t$-weight $(t=3,5,7)$, and then determine their complete weight enumerators by employing  Gauss sums and quadratic character sums. Especially, for $m = 3$, $\mathcal{C}_{D_a}$ is MDS. Furthermore, for $m\ge 4$, the minimum distance of the dual code for above $t$-weight code $(t=3,5)$ is $3$, which implies that these codes are projective. Finally, an $s$-sum set for any odd $s>1$ is constructed based on these three-weight projective codes.
		
	 The paper is organized as follows. In section 2, some related basic notations and results for character sums are given.  In section 3, the complete weight enumerators for several
	 classes of $t$-weight  $(t=3,5,7)$ linear codes are presented,  especially, these $t$-weight $(t=3,5)$ codes are projective. In section 4, the proofs for the main results are given. In section 5, some of these codes can be applicated in secret sharing schemes, and $s$-sum sets for any odd $s>1$ is constructed from this three-weight projective codes. In section 6, we conclude the whole paper.
	 
	\section{Preliminaries}
    \indent An additive character $\chi$ of $\mathbb{F}_{p^m}$ is a function from $\mathbb{F}_{p^m}$ to the multiplicative group $U=\{u\ |\ |u|=1,\ u\in\mathbb{C}\}$, such that $\chi(x+y)=\chi(x)\chi(y)$ for any $x,\ y\in\ \mathbb{F}_{p^m}$. For each $b\in \mathbb{F}_{p^m}$, the function
    \begin{center}
    	$\chi_b(x)=\zeta_p^{\mathrm{Tr}(bx)}$ \quad $(x\in \mathbb{F}_{p^m})$
    \end{center}
    defines an additive character of $\mathbb{F}_{p^m}$. When $b=0,\ \chi_0(x)=1$ for any $x\in \mathbb{F}_{p^m}$ is called the trivial additive character of $\mathbb{F}_{p^m}$. The character $\chi:=\chi_1$ is called the canonical additive character of $\mathbb{F}_{p^m}$ and every
    additive character of $\mathbb{F}_{p^m}$ can be written as $\chi_b(x)=\chi(bx)$. The orthogonal property for the additive character is given by
    \begin{align*}
    	\sum_{x\in \mathbb{F}_{p^m}}\zeta_p^{\mathrm{Tr}(bx)}=\begin{cases}
    		p^m,\quad& \text{if}~b=0;\\
    		0,\quad &\text{otherwise}.
    	\end{cases}
    \end{align*}     
    We extend the quadratic character $\eta_m$ of $\mathbb{F}_{p^m}^*$ by letting $\eta_m(0)=0$, then the quadratic Gauss sums $G_m$ over $\mathbb{F}_{p^m}$ is defined as 
    \begin{center}
    	$G_m=\sum\limits_{x\in \mathbb{F}_{p^m}}\eta_m(x)\chi(x)$.
    \end{center}
    
    \indent     Now, some properties for quadratic character and quadratic Gauss sums are given as follows.    	
    \begin{lemma}[\cite{KD2015}, Lemma 7]\label{l21}
    	For $x\in\mathbb{F}_p^*$, 
    	\begin{align*}
    		\eta_m(x)=\begin{cases}
    			1,\quad& ~2\mid m;\\
    			\eta_{1}(x),\quad &~\text{otherwise}.
    		\end{cases}
    	\end{align*}     	
    \end{lemma} 
    \begin{lemma}[\cite{RL97}, Theorem 5.15]\label{l22}
    	For the Gauss sums $G_m$ over $\mathbb{F}_{p^m}$,
    	\begin{align*}
    		G_m=(-1)^{m-1}\sqrt{-1}^\frac{(p-1)^2m}{4}p^{\frac{m}{2}}.
    	\end{align*}
    \end{lemma}
    
    \begin{lemma}[\cite{RL97}, Theorem 5.33]\label{l23}
    	Let $f(x)=a_2x^2+a_1x+a_0\in \mathbb{F}_{p^m}[x]$ with $a_2\neq0$, then
    	\begin{align*}
    	\sum_{x\in \mathbb{F}_{p^m}}\chi(f(x))=G_m\eta_m(a_2)\chi(a_0-a_1^2(4a_2)^{-1}).
    	\end{align*}
    \end{lemma}

	In order to determine the complete weight enumerator for  $\mathcal{C}_{D_a}$, we need  the following quadratic character sums
	\begin{align*}
	{I}_1(a)=\sum_{z_1\in\mathbb{F}_{p}}\eta_{m}(z_1a+1)\quad\text{and}\quad {I}_{2}(a)=\sum_{z_2\in\mathbb{F}_{p}}\sum_{z_1\in\mathbb{F}_{p}}\eta_{m}(z_2a^2+z_1a+1),
	\end{align*}
	where $a\in\mathbb{F}_{p^m}\backslash\mathbb{F}_p$.

	We give the relation between $I_1(a)$ and $I_{2}(a)$ as follows.
	
	\begin{lemma}\label{l52}
		For any even $m$ and $a\in\mathbb{F}_{p^2}\backslash\mathbb{F}_p$, we have
		\begin{align}\label{l520}
		{I}_{2}(a)=(p-1)\big(\eta_{m}(a)+I_1(a)\big).
		\end{align}
	\end{lemma}	
	
	{\bf Proof.} By $a\in\mathbb{F}_{p^2}\backslash\mathbb{F}_p$, we can assume that there exist some $b_1\in\mathbb{F}_p$ and $b_0\in\mathbb{F}_p^{*}$ such that $a^2+b_1a+b_0=0$, and then by Lemma \ref{l21}, we have
	\begin{align*}\begin{aligned}
	{I}_{2}(a)=&\sum_{z_2\in\mathbb{F}_{p}}\sum_{z_1\in\mathbb{F}_{p}}\eta_{m}(z_2a^2+z_1a+1)\\
	=&\sum_{z_2\in\mathbb{F}_{p}}\sum_{z_1\in\mathbb{F}_{p}}\eta_{m}\big((z_1-b_1z_2)a+(1-b_0z_2)\big)\\
	=&\sum_{z_1\in\mathbb{F}_p}\eta_{m}\big((z_1-b_1b_0^{-1})a\big)+\sum_{z_2\in\mathbb{F}_{p}\backslash\{b_0^{-1}\}}\sum_{z_1\in\mathbb{F}_{p}}\eta_{m}\big((z_1-b_1z_2)a+(1-b_0z_2)\big)\\
	=&\eta_{m}(a)\sum_{z_1\in\mathbb{F}_p^{*}}\eta_{m}(z_1)+\sum_{z_2\in\mathbb{F}_{p}\backslash\{b_0^{-1}\}}\sum_{z_1\in\mathbb{F}_{p}}\eta_{m}\big(z_1a+(1-b_0z_2)\big)\\
	=&(p-1)\eta_{m}(a)+\sum_{z_2\in\mathbb{F}_{p}\backslash\{b_0^{-1}\}}\sum_{z_1\in\mathbb{F}_{p}}\eta_{m}\big(z_1a+1\big)\\
	=&(p-1)\big(\eta_{m}(a)+I_1(a)\big).
	\end{aligned}
	\end{align*} $\hfill\Box$
	
In the following two lemmas, for some even $m$ and $a\in\mathbb{F}_{p^{m}}\backslash\mathbb{F}_p$, the explicit value of $I_i(a)$ $(i=1,2)$ is determined.

	
	
	
	
	\begin{lemma}\label{l50}
		For any integer $s$ and any even $m$, the following assertions hold.
		
		$(1)$ If $s>2$, $s\mid \frac{m}{2}$ and  $a\in\mathbb{F}_{p^s}\backslash\mathbb{F}_{p^2}$, then
		\begin{align}\label{l501}
		I_1(a)=p\quad\text{and}\quad I_{2}(a)=p^2.
		\end{align}
		
		$(2)$ If $4\mid m$ and $a\in\mathbb{F}_{p^2}\backslash\mathbb{F}_{p}$, then
		\begin{align}\label{l511}
		I_1(a)=p\quad\text{and}\quad I_{2}(a)=p^2-1.
		\end{align}
	\end{lemma}
	
	{\bf Proof.} 
	$(1)$ By the assumptions that $s\mid \frac{m}{2}$ and $a\in\mathbb{F}_{p^s}\backslash\mathbb{F}_{p^2}$, we know that for any $(z_1, z_2)\in\mathbb{F}_p^2$, $z_1a+1$ and $z_2a^2+z_1a+1$ are both square elements in $\mathbb{F}_{p^m}^{*}$, thus $(\ref{l501})$ holds.

	$(2)$ By the assumptions that $4\mid m$ and $a\in\mathbb{F}_{p^2}\backslash\mathbb{F}_p$, we know that for any $(z_1, z_2)\in\mathbb{F}_p^2$, $z_1a+1$ and $z_2a^2+z_1a+1$ are both square elements in $\mathbb{F}_{p^m}^{*}$.
	 Note that $a\in\mathbb{F}_{p^2}\backslash\mathbb{F}_p$, we can get $z_1a+1\neq 0$, thus $I_1(a)=p$.
	 Furthermore, we can assume that there exist some $b_1\in\mathbb{F}_p$ and $b_0\in\mathbb{F}_p^{*}$ such that $a^2+b_1a+b_0=0$, and then 
	$z_2a^2+z_1a+1=0$ if and only if $z_2=b_0^{-1}$ and $z_1=b_0^{-1}b_1$, thus  $I_{2}(a)=p^2-1$.
	$\hfill\Box$\\


	The following Pless power moments are useful for calculating the minimum distance of the dual code for a
	linear code.
	

	\begin{lemma}[\cite{WC2003}, p.259, The Pless power moments]\label{l12}
		For an $[n, k, d]$ code $\mathcal{C}$ over $\mathbb{F}_p$ with the weight distribution $(1, A_1,\ldots, A_n)$, suppose that
		the weight distribution  of  its dual code is $(1, A_1^{\bot},\ldots,A_n^{\bot})$, then the first four Pless power
		moments are
		\begin{align*}
	&\sum_{j=0}^{n}A_j=p^{k},\\
&\sum_{j=0}^{n}jA_j=p^{k-1}\Big(pn-n- A_1^{\bot}\Big),\\
&		\sum_{j=0}^{n}j^2A_j=p^{k-2}\Big((p-1)n(pn-n+1)-(2pn-p-2n+2)A_1^{\bot}+2A_2^{\bot}\Big),\\
&		\sum_{j=0}^{n}j^3A_j=p^{k-3}\Big((p-1)n(p^2n^2-2pn^2+3pn-p+n^2-3n+2)\\
		&\qquad\qquad\qquad\quad~~-(3p^2n^2-3p^2n-6pn^2+12pn+p^2-6p+3n^2-9n+6)A_1^{\bot}\\
		&\qquad\qquad\qquad\quad~~+6(pn-p-n+2)A_2^{\bot}-6A_3^{\bot}\Big),
		\end{align*}
	respectively.
	\end{lemma}
    \section{Main Results}
    
    \begin{theorem}\label{t1}
    	For any odd $m\ge 3$  and $a\in\mathbb{F}_{p^{m}}\backslash\mathbb{F}_p$,  $\mathcal{C}_{D_a}$ is a $\big{[}p^{m-2}, m, p^{m-2}-p^{m-3}-p^{\frac{m-3}{2}}\big{]}$ code with the weight distribution in Table $1$, and the complete weight enumerator is {\small	\begin{align}\label{t11}\begin{aligned}
    		W(\mathcal{C}_{D_a})=&1+\big(p^{m-1}-p^{m-2}+p^{m-3}-1\big)\prod_{\rho\in \mathbb{F}_p}w_{\rho}^{p^{m-3}}\\
    		+&\frac{1}{2}(p-1)p^{m-2}w_0^{p^{m-3}}\prod_{\rho\in \mathbb{F}_p^{*}}w_{\rho}^{p^{m-3}- \eta_{1}(-\rho)p^{\frac{m-3}{2}}}
    		+\frac{1}{2}(p-1)p^{m-2}w_0^{p^{m-3}}\prod_{\rho\in \mathbb{F}_p^{*}}w_{\rho}^{p^{m-3}+ \eta_{1}(-\rho)p^{\frac{m-3}{2}}}\\
    		+&\frac{1}{2}\big(p^{m-3}-I_{2}(a)p^{\frac{m-5}{2}}\big)\sum_{\gamma\in\mathbb{F}_p^{*}}w_\gamma^{p^{m-3}-(p-1) p^{\frac{m-3}{2}}}\prod_{\rho\in \mathbb{F}_p\backslash\{\gamma\}}w_{\rho}^{p^{m-3}+p^{\frac{m-3}{2}}}\\
    		+&\frac{1}{2}\big(p^{m-3}+I_{2}(a)p^{\frac{m-5}{2}}\big) \sum_{\gamma\in\mathbb{F}_p^{*}}w_\gamma^{	p^{m-3}+(p-1) p^{\frac{m-3}{2}}}\prod_{\rho\in \mathbb{F}_p\backslash\{\gamma\}}w_{\rho}^{p^{m-3}-p^{\frac{m-3}{2}}}\\	
    		+&\frac{1}{2}\Big((p-1)p^{m-2}+\eta_{1}(-1) \big(I_{2}(a)-p\big)p^{\frac{m-5}{2}}\Big)\sum_{\gamma\in\mathbb{F}_p^{*}}w_\gamma^{p^{m-3}}\prod_{\rho\in \mathbb{F}_p\backslash\{\gamma\}}w_{\rho}^{p^{m-3}+\eta_{1}(\rho\gamma^{-1}-1)p^{\frac{m-3}{2}}}\\
    		+&\frac{1}{2}\Big((p-1)p^{m-2}-\eta_{1}(-1) \big(I_{2}(a)-p\big)p^{\frac{m-5}{2}}\Big)\sum_{\gamma\in\mathbb{F}_p^{*}}w_\gamma^{p^{m-3}}\prod_{\rho\in \mathbb{F}_p\backslash\{\gamma\}}w_{\rho}^{p^{m-3}-\eta_{1}(\rho\gamma^{-1}-1)p^{\frac{m-3}{2}}}.
    		\end{aligned}\end{align}}\\
    	
    	\begin{center} Table $1$~~~ The weight distribution for  $\mathcal{C}_{D_a}$ $(m\text{~is~odd},~a\in\mathbb{F}_{p^{m}}\backslash\mathbb{F}_p)$	
    		\begin{tabular}{|p{4cm}<{\centering}| p{8cm}<{\centering}|}
    			\hline   weight $w$	                     &   frequency $A_w$                    \\ 
    			\hline       $0$	                     &  $1$                                    \\ 
    			\hline   $ p^{m-2}-p^{m-3}-p^{\frac{m-3}{2}}$   &  $\frac{1}{2}(p-1)\big((p-1)p^{m-2}+p^{m-3}-p^\frac{m-3}{2}\big)\big)$        \\ 
    			\hline   $ p^{m-2}-p^{m-3}$   &  $2(p-1)p^{m-2}+p^{m-3}-1$        \\ 
    			\hline   $ p^{m-2}-p^{m-3}+p^{\frac{m-3}{2}}$&       $\frac{1}{2}(p-1)\big((p-1)p^{m-2}+p^{m-3}+p^\frac{m-3}{2}\big)$    \\         
    			\hline
    		\end{tabular}
    	\end{center} 
    \end{theorem}	
	\begin{theorem}\label{t2}
		For any even $m\ge 4$  and $a\in\mathbb{F}_{p^{2}}\backslash\mathbb{F}_{p}$, $\mathcal{C}_{D_a}$ is a $\big{[}p^{m-2}, m, p^{m-2}-p^{m-3}-p^{\frac{m-4}{2}}\big{]}$ code with the weight distribution in Table $2$, and the complete weight enumerator is{\small	\begin{align*}
			W(\mathcal{C}_{D_a})=&1+\big(p^{m-2}-1\big)\prod_{\rho\in \mathbb{F}_p}w_{\rho}^{p^{m-3}}\\
			&+	\frac{1}{2}\Big(p^{m-1}+p^{\frac{m-4}{2}}\big(p^2-I_{2}(a)\big)\Big)\sum_{\gamma\in\mathbb{F}_p^{*}}w_\gamma^{
				p^{m-3}+(p-1)p^{\frac{m-4}{2}}}\prod_{\rho\in \mathbb{F}_p\backslash\{\gamma\}}w_{\rho}^{p^{m-3}-p^{\frac{m-4}{2}}}\\
			&+	\frac{1}{2}\Big(p^{m-1}- p^{\frac{m-4}{2}}\big(p^2-I_{2}(a)\big)\Big)\sum_{\gamma\in\mathbb{F}_p^{*}}w_\gamma^{
				p^{m-3}-(p-1)p^{\frac{m-4}{2}}}\prod_{\rho\in \mathbb{F}_p\backslash\{\gamma\}}w_{\rho}^{p^{m-3}+p^{\frac{m-4}{2}}}\\
			&+	\frac{1}{2}(p-1)\big(p^{m-2}+ p^{\frac{m-4}{2}}\big((p-1)-\eta_{m}(a)I_1(a)\big)\big)w_0^{
				p^{m-3}+(p-1)p^{\frac{m-4}{2}}}\prod_{\rho\in \mathbb{F}_p^*}w_{\rho}^{p^{m-3}-p^{\frac{m-4}{2}}}\\
			&+	\frac{1}{2}(p-1)\big(p^{m-2}-p^{\frac{m-4}{2}}\big((p-1)-\eta_{m}(a)I_1(a)\big)\big)w_0^{
				p^{m-3}-(p-1)p^{\frac{m-4}{2}}}\prod_{\rho\in \mathbb{F}_p^*}w_{\rho}^{p^{m-3}+p^{\frac{m-4}{2}}}.
			\end{align*}}	
		\begin{center} Table $2$~~~ The weight distribution for  $\mathcal{C}_{D_a}$ $(2\mid m, ~a\in\mathbb{F}_{p^{2}}\backslash\mathbb{F}_p)$
		\begin{tabular}{|p{5cm}<{\centering}| p{8cm}<{\centering}|}
			\hline   weight $w$	                     &   frequency $A_w$                    \\ 
			\hline       $0$	                     &  $1$                                    \\
			\hline   $ p^{m-2}-p^{m-3}-(p-1)p^{\frac{m-4}{2}}$   & 	$\frac{1}{2}(p-1)\big(p^{m-2}+p^{\frac{m-4}{2}}\big((p-1)-\eta_{m}(a)I_1(a)\big)\big)$         \\ 
			\hline   $ p^{m-2}-p^{m-3}-p^{\frac{m-4}{2}}$   &  $\frac{1}{2}(p-1)\big(p^{m-1}-p^{\frac{m-4}{2}}\big(p^2-I_{2}(a)\big)\big) $        \\ 
			\hline   $ p^{m-2}-p^{m-3}$   &  $p^{m-2}-1$        \\ 
			\hline   $ p^{m-2}-p^{m-3}+p^{\frac{m-4}{2}}$   &  $\frac{1}{2}(p-1)\big(p^{m-1}+p^{\frac{m-4}{2}}\big(p^2-I_{2}(a)\big)\big) $        \\
			\hline   $ p^{m-2}-p^{m-3}+(p-1)p^{\frac{m-4}{2}}$   & 	$\frac{1}{2}(p-1)\big(p^{m-2}-p^{\frac{m-4}{2}}\big((p-1)-\eta_{m}(a)I_1(a)\big)\big)$         \\ 
			\hline
		\end{tabular}
	\end{center} 
\end{theorem}
	\begin{theorem}\label{t3}
	For any even $m\ge 4$  and $a\in\mathbb{F}_{p^{m}}\backslash\mathbb{F}_{p^2}$,  $\mathcal{C}_{D_a}$ is a $\big{[}p^{m-2}, m, p^{m-2}-p^{m-3}-p^{\frac{m-2}{2}}\big{]}$ code with the weight distribution in Table $3$, and the complete weight enumerator is  
	{\small	\begin{align*}
		W(\mathcal{C}_{D_a})=&1+\big(p^{m-1}-p^{m-2}+p^{m-3}-1\big)\prod_{\rho\in \mathbb{F}_p}w_{\rho}^{p^{m-3}}\\
		&+\frac{1}{2}\Big(p^{m-3}- p^{\frac{m-6}{2}}\big((p-1)\big(1+\eta_{m}(a)I_1(a)\big)-I_{2}(a)\big)\Big)\sum_{\gamma\in\mathbb{F}_p^{*}}w_\gamma^{p^{m-3}}\prod_{\rho\in \mathbb{F}_p\backslash\{\gamma\}}w_{\rho}^{p^{m-3}-\eta_{1}(1-\rho\gamma^{-1})p^{\frac{m-2}{2}}}\\
		&+\frac{1}{2}\Big(p^{m-3}+ p^{\frac{m-6}{2}}\big((p-1)\big(1+\eta_{m}(a)I_1(a)\big)-I_{2}(a)\big)\Big)\sum_{\gamma\in\mathbb{F}_p^{*}}w_\gamma^{p^{m-3}}\prod_{\rho\in \mathbb{F}_p\backslash\{\gamma\}}w_{\rho}^{p^{m-3}+\cdot\eta_{1}(1-\rho\gamma^{-1})p^{\frac{m-2}{2}}}\\
		&+	\frac{1}{2}(p-1)\Big(p^{m-2}+p^{\frac{m-4}{2}}\big((p-1)-\eta_{m}(a)I_1(a)\big)\Big)w_0^{p^{m-3}+(p-1)p^{\frac{m-4}{2}}}\prod_{\rho\in \mathbb{F}_p^{*}}w_{\rho}^{p^{m-3}- p^{\frac{m-4}{2}}}\\
		&+	\frac{1}{2}(p-1)\Big(p^{m-2}-p^{\frac{m-4}{2}}\big((p-1)-\eta_{m}(a)I_1(a)\big)\Big)w_0^{p^{m-3}-(p-1)p^{\frac{m-4}{2}}}\prod_{\rho\in \mathbb{F}_p^{*}}w_{\rho}^{p^{m-3}+p^{\frac{m-4}{2}}}\\
		&+\frac{1}{2}\Big((p-1)p^{m-2}+p^{\frac{m-4}{2}}\big(p^2-I_{2}(a)\big)\Big)\sum_{\gamma\in\mathbb{F}_p^{*}}w_\gamma^{
			p^{m-3}+(p-1)p^{\frac{m-4}{2}}}\prod_{\rho\in \mathbb{F}_p\backslash\{\gamma\}}w_{\rho}^{p^{m-3}-p^{\frac{m-4}{2}}}\\
		&+\frac{1}{2}\Big((p-1)p^{m-2}-p^{\frac{m-4}{2}}\big(p^2-I_{2}(a)\big)\Big)\sum_{\gamma\in\mathbb{F}_p^{*}}w_\gamma^{
			p^{m-3}-(p-1)p^{\frac{m-4}{2}}}\prod_{\rho\in \mathbb{F}_p\backslash\{\gamma\}}w_{\rho}^{p^{m-3}+p^{\frac{m-4}{2}}}.
		\end{align*}}
	\begin{center} Table $3$~~~ The weight distribution for  $\mathcal{C}_{D_a}$ $(2\mid m,~ a\in\mathbb{F}_{p^{m}}\backslash\mathbb{F}_{p^2})$
		{\small
		\begin{tabular}{|p{6cm}<{\centering}| p{10cm}<{\centering}|}
			\hline   weight $w$	                     &   frequency $A_w$                    \\ 
			\hline       $0$	                     &  $1$                                    \\ 
			\hline   $ p^{m-2}-p^{m-3}-p^{\frac{m-2}{2}}$   &  $\frac{1}{2}(p-1)\Big(p^{m-3}+p^{\frac{m-6}{2}}\big((p-1)\big(1+\eta_{m}(a)I_1(a)\big)-I_{2}(a)\big)\Big)$        \\ 
			\hline   $ p^{m-2}-p^{m-3}-(p-1)p^{\frac{m-4}{2}}$   &  $
			\frac{1}{2}(p-1)\Big(p^{m-2}+p^{\frac{m-4}{2}}\big((p-1)-\eta_{m}(a)I_1(a)\big)\Big)$        \\ 
			\hline   $ p^{m-2}-p^{m-3}-p^{\frac{m-4}{2}}$   & $	\frac{1}{2}(p-1)\big((p-1)p^{m-2}-p^{\frac{m-4}{2}}\big(p^2-I_{2}(a)\big)\big)$         \\ 
			\hline   $ p^{m-2}-p^{m-3}$   &  $p^{m-1}-p^{m-2}+p^{m-3}-1$        \\ 
			\hline   $ p^{m-2}-p^{m-3}+p^{\frac{m-4}{2}}$   &  $	\frac{1}{2}(p-1)\big((p-1)p^{m-2}+p^{\frac{m-4}{2}}\big(p^2-I_{2}(a)\big)\big)$          \\ 
			\hline   $ p^{m-2}-p^{m-3}+(p-1)p^{\frac{m-4}{2}}$   &  $
			\frac{1}{2}(p-1)\Big(p^{m-2}-p^{\frac{m-4}{2}}\big((p-1)-\eta_{m}(a)I_1(a)\big)\Big)$       \\ 	\hline   $ p^{m-2}-p^{m-3}+p^{\frac{m-2}{2}}$   &  $\frac{1}{2}(p-1)\Big(p^{m-3}-p^{\frac{m-6}{2}}\big((p-1)\big(1+\eta_{m}(a)I_1(a)\big)-I_{2}(a)\big)\Big)$        \\ 
			\hline
		\end{tabular}} 
	\end{center} 
	\end{theorem}

\begin{remark}
	$(1)$ For any odd $m$, the weight distribution for $\mathcal{C}_{D_a}$ is determined explictly, but the complete weight enumerator is depended on the value of  $I_2(a)$.
	
	$(2)$ For any even $m$, both the  weight distribution and the complete weight enumerator for $\mathcal{C}_{D_a}$  are depended on the value of  $I_i(a)$ $(i=1,2)$.
\end{remark}

 
In the following corollary, we give the complete weight enumerator for an MDS code with parametres $[p,3,p-2]$, explicitly.
\begin{corollary}\label{c31}
For  $m=3$  and $a\!\in\!\mathbb{F}_{p^{3}}\backslash\mathbb{F}_p$,  $\mathcal{C}_{D_a}$ is an MDS code with parameters\! $\big{[}p, 3, p-2\big{]}$, and 
the complete weight enumerator is {\small	\begin{align}\label{c10}\begin{aligned}		
	W(\mathcal{C}_{D_a})=&1+p(p-1)\prod_{\rho\in \mathbb{F}_p}w_{\rho}+\frac{p(p-1)}{2}w_0\prod_{\rho\in \mathbb{F}_p^{*}}w_{\rho}^{1-\eta_{1}(-\rho)}
	+\frac{p(p-1)}{2}w_0\prod_{\rho\in \mathbb{F}_p^{*}}w_{\rho}^{1+\eta_{1}(-\rho)}+\sum_{\gamma\in\mathbb{F}_p^{*}}w_\gamma^{p}\\
	&
	+\frac{p(p-1)}{2}\sum_{\gamma\in\mathbb{F}_p^{*}}w_\gamma\!\!\prod_{\rho\in \mathbb{F}_p\backslash\{\gamma\}}\!\!w_{\rho}^{1-\eta_{1}(\rho\gamma^{-1}-1)}
	+\frac{p(p-1)}{2}\sum_{\gamma\in\mathbb{F}_p^{*}}w_\gamma\!\!\prod_{\rho\in \mathbb{F}_p\backslash\{\gamma\}}\!\!w_{\rho}^{1+\eta_{1}(\rho\gamma^{-1}-1)}.	\end{aligned}
	\end{align}}
\end{corollary}


{\bf Proof.} 
By taking $m=3$ in Theorem \ref{t1}, we know that $\mathcal{C}_{D_a}$ is an MDS code with parameters $\big{[}p, 3, p-2\big{]}$, and the complete weight enumerator is {\small	\begin{align}\label{c12}\begin{aligned}
W(\mathcal{C}_{D_a})=&1+p(p-1)\prod_{\rho\in \mathbb{F}_p}w_{\rho}+\frac{p(p-1)}{2}w_0\prod_{\rho\in \mathbb{F}_p^{*}}w_{\rho}^{1-\eta_{1}(-\rho)}
+\frac{p(p-1)}{2}w_0\prod_{\rho\in \mathbb{F}_p^{*}}w_{\rho}^{1+\eta_{1}(-\rho)}\\
&+\frac{1-I_{2}(a)\frac{1}{p}}{2}\sum_{\gamma\in\mathbb{F}_p^{*}}w_\gamma^{2-p}\prod_{\rho\in \mathbb{F}_p\backslash\{\gamma\}}w_{\rho}^{2}+\frac{1+I_{2}(a)\frac{1}{p}}{2}\sum_{\gamma\in\mathbb{F}_p^{*}}w_\gamma^{p}\\	
&+\frac{(p-1)p+\eta_{1}(-1) \big(I_{2}(a)-p\big)\frac{1}{p}}{2}\sum_{\gamma\in\mathbb{F}_p^{*}}w_\gamma\prod_{\rho\in \mathbb{F}_p\backslash\{\gamma\}}w_{\rho}^{1+\eta_{1}(\rho\gamma^{-1}-1)}\\
&+\frac{(p-1)p-\eta_{1}(-1) \big(I_{2}(a)-p\big)\frac{1}{p}}{2}\sum_{\gamma\in\mathbb{F}_p^{*}}w_\gamma\prod_{\rho\in \mathbb{F}_p\backslash\{\gamma\}}w_{\rho}^{1-\eta_{1}(\rho\gamma^{-1}-1)}.
\end{aligned}
\end{align}

Note that $2-p<0$, thus there is no any codeword in $\mathcal{C}_{D_a}$ with the complete weight enumerator $w_\gamma^{2-p}\prod\limits_{\rho\in \mathbb{F}_p\backslash\{\gamma\}}w_{\rho}^{2}$, 
it implies that $\frac{1-I_{2}(a)\frac{1}{p}}{2}=0$, and then
\begin{align}\label{c13}
I_{2}(a)=p.
\end{align}
Now by $(\ref{c12})$-$(\ref{c13})$, we can get $(\ref{c10})$.$\hfill\Box$\\

\indent By Lemma \ref{l50}, Remark $\ref{r1}$ and Theorems \ref{t2}-\ref{t3}, we can present the weight distributions 
and the complete weight enumerators explicitly for some cases in Theorems $\ref{c2}$-$\ref{c4}$.
\begin{theorem}\label{c2}
		For any integer $m$ with $4 \mid m$, if $a\! \in\! \mathbb{F}_{p^{2}}\backslash\mathbb{F}_{p}$, then $\mathcal{C}_{D_a}$ is a $[p^{m-2}, m, p^{m-2}\!-\!p^{m-3}\!-\!(\!p-1\!)p^{\frac{m-4}{2}}]$ code with the weight distribution in Table $4$, and the complete weight enumerator is
	{\small
	\begin{align*}
		W(\mathcal{C}_{D_a})=&1+\big(p^{m-2}-1\big)\prod_{\rho\in \mathbb{F}_p}w_{\rho}^{p^{m-3}}\\
		&+	\frac{1}{2}\big(p^{m-1}+p^{\frac{m-4}{2}}\big) \sum_{\gamma\in\mathbb{F}_p^{*}}w_\gamma^{
			p^{m-3}+(p-1)p^{\frac{m-4}{2}}}\prod_{\rho\in \mathbb{F}_p\backslash\{\gamma\}}w_{\rho}^{p^{m-3}-p^{\frac{m-4}{2}}}\\
		&+	\frac{1}{2}\big(p^{m-1}- p^{\frac{m-4}{2}}\big)\sum_{\gamma\in\mathbb{F}_p^{*}}w_\gamma^{
			p^{m-3}-(p-1)p^{\frac{m-4}{2}}}\prod_{\rho\in \mathbb{F}_p\backslash\{\gamma\}}w_{\rho}^{p^{m-3}+p^{\frac{m-4}{2}}}\\
		&+	\frac{1}{2}(p-1)\big(p^{m-2}-p^{\frac{m-4}{2}}\big)w_0^{
			p^{m-3}+(p-1)p^{\frac{m-4}{2}}}\prod_{\rho\in \mathbb{F}_p^*}w_{\rho}^{p^{m-3}-p^{\frac{m-4}{2}}}\\
		&+	\frac{1}{2}(p-1)\big(p^{m-2}+p^{\frac{m-4}{2}}\big)w_0^{
			p^{m-3}-(p-1)p^{\frac{m-4}{2}}}\prod_{\rho\in \mathbb{F}_p^*}w_{\rho}^{p^{m-3}+p^{\frac{m-4}{2}}}.
		\end{align*}
	}
	\begin{center} Table $4$~~~ The weight distribution for  $\mathcal{C}_{D_a}$ $(4\mid m,~a\in\mathbb{F}_{p^{2}}\backslash\mathbb{F}_p)$
		\begin{tabular}{|p{6cm}<{\centering}| p{6cm}<{\centering}|}
			\hline   weight $w$	                     &   frequency $A_w$                    \\ 
			\hline       $0$	                     &  $1$                                    \\ 
			\hline   $ p^{m-2}-p^{m-3}-(p-1)p^{\frac{m-4}{2}}$   & 	$\frac{1}{2}(p-1)\big(p^{m-2}-p^{\frac{m-4}{2}}\big)$         \\ 
			\hline   $ p^{m-2}-p^{m-3}-p^{\frac{m-4}{2}}$   &  $\frac{1}{2}(p-1)\big(p^{m-1}-p^{\frac{m-4}{2}}\big) $        \\ 
			\hline   $ p^{m-2}-p^{m-3}$   &  $p^{m-2}-1$        \\ 
			\hline   $ p^{m-2}-p^{m-3}+p^{\frac{m-4}{2}}$   &  $\frac{1}{2}(p-1)\big(p^{m-1}+p^{\frac{m-4}{2}}\big) $        \\
			\hline   $ p^{m-2}-p^{m-3}+(p-1)p^{\frac{m-4}{2}}$   & 	$\frac{1}{2}(p-1)\big(p^{m-2}+p^{\frac{m-4}{2}}\big)$         \\ 
			\hline
		\end{tabular} 
	\end{center} 
\end{theorem}
\begin{theorem}\label{c3}
		For any integer $s$ and even $m$ with $s>2$ and $s\!\mid\! \frac{m}{2}$, if $a\!\in\!\mathbb{F}_{p^{s}}\backslash\mathbb{F}_{p^2}$, then $\mathcal{C}_{D_a}$ is a {\small$[p^{m-2}, m, p^{m-2}-p^{m-3}-p^{\frac{m-2}{2}}]$} code with the weight distribution in Table $5$, and the complete weight enumerator is  {\small	\begin{align*}
		W(\mathcal{C}_{D_a})=&1+\big(p^{m-1}-p^{m-2}+p^{m-3}-1\big)\prod_{\rho\in \mathbb{F}_p}w_{\rho}^{p^{m-3}}\\
		&+\frac{1}{2}\big(p^{m-3}+p^{\frac{m-6}{2}}\big)\sum_{\gamma\in\mathbb{F}_p^{*}}w_\gamma^{p^{m-3}}\prod_{\rho\in \mathbb{F}_p\backslash\{\gamma\}}w_{\rho}^{p^{m-3}-\eta_{1}(1-\rho\gamma^{-1})p^{\frac{m-2}{2}}}\\
		&+\frac{1}{2}\big(p^{m-3}- p^{\frac{m-6}{2}}\big)\sum_{\gamma\in\mathbb{F}_p^{*}}w_\gamma^{p^{m-3}}\prod_{\rho\in \mathbb{F}_p\backslash\{\gamma\}}w_{\rho}^{p^{m-3}+\eta_{1}(1-\rho\gamma^{-1})p^{\frac{m-2}{2}}}\\
		&+	\frac{1}{2}(p-1)\big(p^{m-2}-p^{\frac{m-4}{2}}\big)w_0^{p^{m-3}+(p-1)p^{\frac{m-4}{2}}}\prod_{\rho\in \mathbb{F}_p^{*}}w_{\rho}^{p^{m-3}- p^{\frac{m-4}{2}}}\\
		&+	\frac{1}{2}(p-1)\big(p^{m-2}+p^{\frac{m-4}{2}}\big)w_0^{p^{m-3}-(p-1)p^{\frac{m-4}{2}}}\prod_{\rho\in \mathbb{F}_p^{*}}w_{\rho}^{p^{m-3}+p^{\frac{m-4}{2}}}\\
		&+\frac{1}{2}(p-1)p^{m-2}\sum_{\gamma\in\mathbb{F}_p^{*}}w_\gamma^{
			p^{m-3}+(p-1)p^{\frac{m-4}{2}}}\prod_{\rho\in \mathbb{F}_p\backslash\{\gamma\}}w_{\rho}^{p^{m-3}-p^{\frac{m-4}{2}}}\\
		&+\frac{1}{2}(p-1)p^{m-2}\sum_{\gamma\in\mathbb{F}_p^{*}}w_\gamma^{
			p^{m-3}-(p-1)p^{\frac{m-4}{2}}}\prod_{\rho\in \mathbb{F}_p\backslash\{\gamma\}}w_{\rho}^{p^{m-3}+p^{\frac{m-4}{2}}}.
		\end{align*}}
	
	\begin{center} Table $5$~~~ The weight distribution for $\mathcal{C}_{D_a}$  $(s\mid\frac{m}{2},~a\in\mathbb{F}_{p^{s}}\backslash\mathbb{F}_{p^2})$			
	{\small	\begin{tabular}{|p{6cm}<{\centering}| p{6cm}<{\centering}|}
			\hline   weight $w$	                     &   frequency $A_w$                    \\ 
			\hline       $0$	                     &  $1$                                    \\ 
			\hline   $ p^{m-2}-p^{m-3}-p^{\frac{m-2}{2}}$   &  $\frac{1}{2}(p-1)\big(p^{m-3}-p^{\frac{m-6}{2}}\big)$        \\ 
			\hline   $ p^{m-2}-p^{m-3}-(p-1)p^{\frac{m-4}{2}}$   &  $\frac{1}{2}(p-1)\big(p^{m-2}-p^{\frac{m-4}{2}}\big)$        \\ 
			\hline   $ p^{m-2}-p^{m-3}-p^{\frac{m-4}{2}}$   & $	\frac{1}{2}(p-1)^2p^{m-2}$         \\ 
			\hline   $ p^{m-2}-p^{m-3}$   &  $p^{m-1}-p^{m-2}+p^{m-3}-1$        \\ 
			\hline   $ p^{m-2}-p^{m-3}+p^{\frac{m-4}{2}}$   &  $	\frac{1}{2}(p-1)^2p^{m-2}$          \\ 
			\hline   $ p^{m-2}-p^{m-3}+(p-1)p^{\frac{m-4}{2}}$   &  $\frac{1}{2}(p-1)\big(p^{m-2}+p^{\frac{m-4}{2}}\big)$      \\ 	
			\hline   $ p^{m-2}-p^{m-3}+p^{\frac{m-2}{2}}$   &  $\frac{1}{2}(p-1)\big(p^{m-3}+p^{\frac{m-6}{2}}\big)$        \\ 
			\hline
		\end{tabular} }
	\end{center} 
\end{theorem}	\begin{theorem}\label{c4}
	For any even $m\ge 4$, if $a\in\mathbb{F}_{p^{2}}\backslash\mathbb{F}_{p}$ with $\eta_{m}(a)=-1$, then $\mathcal{C}_{D_a}$ is a $\big{[}p^{m-2}, m, p^{m-2}-p^{m-3}-(p-1)p^{\frac{m-4}{2}}\big{]}$ code with the weight distribution in Table $6$, and the complete weight enumerator is{\small	\begin{align*}
		W(\mathcal{C}_{D_a})=&1+\big(p^{m-2}-1\big)\prod_{\rho\in \mathbb{F}_p}w_{\rho}^{p^{m-3}}\\
		&+	\frac{1}{2}\big(p^{m-1}+p^{\frac{m}{2}}\big)\sum_{\gamma\in\mathbb{F}_p^{*}}w_\gamma^{
			p^{m-3}+(p-1)p^{\frac{m-4}{2}}}\prod_{\rho\in \mathbb{F}_p\backslash\{\gamma\}}w_{\rho}^{p^{m-3}-p^{\frac{m-4}{2}}}\\
		&+	\frac{1}{2}\big(p^{m-1}- p^{\frac{m}{2}}\big)\sum_{\gamma\in\mathbb{F}_p^{*}}w_\gamma^{
			p^{m-3}-(p-1)p^{\frac{m-4}{2}}}\prod_{\rho\in \mathbb{F}_p\backslash\{\gamma\}}w_{\rho}^{p^{m-3}+p^{\frac{m-4}{2}}}\\
		&+	\frac{1}{2}(p-1)\big(p^{m-2}+ p^{\frac{m-2}{2}}\big)w_0^{
			p^{m-3}+(p-1)p^{\frac{m-4}{2}}}\prod_{\rho\in \mathbb{F}_p^*}w_{\rho}^{p^{m-3}-p^{\frac{m-4}{2}}}\\
		&+	\frac{1}{2}(p-1)\big(p^{m-2}-p^{\frac{m-2}{2}}\big)w_0^{
			p^{m-3}-(p-1)p^{\frac{m-4}{2}}}\prod_{\rho\in \mathbb{F}_p^*}w_{\rho}^{p^{m-3}+p^{\frac{m-4}{2}}}.
		\end{align*}}	
	\begin{center} Table $6$~~~ The weight distribution for   $\mathcal{C}_{D_a}$ $(2\mid m,~ a\in\mathbb{F}_{p^{2}}\backslash\mathbb{F}_p,~\eta_{m}(a)=-1)$
		\begin{tabular}{|p{6cm}<{\centering}| p{6cm}<{\centering}|}
			\hline   weight $w$	                     &   frequency $A_w$                    \\ 
			\hline       $0$	                     &  $1$                                    \\ 
			\hline   $ p^{m-2}-p^{m-3}-(p-1)p^{\frac{m-4}{2}}$   & 	$\frac{1}{2}(p-1)\big(p^{m-2}+p^{\frac{m-2}{2}}\big)$         \\ 
			\hline   $ p^{m-2}-p^{m-3}-p^{\frac{m-4}{2}}$   &  $\frac{1}{2}(p-1)\big(p^{m-1}-p^{\frac{m}{2}}\big) $        \\ 
			\hline   $ p^{m-2}-p^{m-3}$   &  $p^{m-2}-1$        \\ 
			\hline   $ p^{m-2}-p^{m-3}+p^{\frac{m-4}{2}}$   &  $\frac{1}{2}(p-1)\big(p^{m-1}+p^{\frac{m}{2}}\big) $        \\
			\hline   $ p^{m-2}-p^{m-3}+(p-1)p^{\frac{m-4}{2}}$   & 	$\frac{1}{2}(p-1)\big(p^{m-2}-p^{\frac{m-2}{2}}\big)$         \\ 
			\hline
		\end{tabular} 
	\end{center} 
\end{theorem}	

In the following theorem, the parameters for  $\mathcal{C}_{D_a}^{\perp}$ are given. 

\begin{theorem}\label{c5}
	For any integers $s>2$ and $m\ge4$, the following assertions hold.\\
	
	$(1)$ If $m$ is odd and $a\in\mathbb{F}_{p^m}\backslash\mathbb{F}_p$, or  $4\mid m$  and $a\in\mathbb{F}_{p^{2}}\backslash\mathbb{F}_{p}$, or $m$ is even  and $a\in\mathbb{F}_{p^{2}}\backslash\mathbb{F}_{p}$ with $\eta_{m}(a)=-1$, then  $C_{D_a}^{\perp}$ is a $\big{[}p^{m-2},p^{m-2}-m,3\big{]}$ linear code, namely $C_{D_a}$ is projective.\\
	
	$(2)$ If $m$ is even with $s\mid \frac{m}{2}$ and $a\in\mathbb{F}_{p^{s}}\backslash\mathbb{F}_{p^2}$, then  $C_{D_a}^{\perp}$ is a  $\big{[}p^{m-2},p^{m-2}-m,2\big{]}$ linear code.
\end{theorem}

{\bf Proof.} Let $A_i^{\perp}\ (i=1,\ldots,n)$ be the number of nonzero codewords with Hamming weight $i$ in $\mathcal{C}_{D_a}^{\perp}$. For $m\ge4$, by Lemma \ref{l12}, Theorems $\ref{t1}$ and $\ref{c2}$-$\ref{c4}$,  we have the following four cases.

{\bf Case 1.}  If $m$ is odd and $a\in\mathbb{F}_{p^m}\backslash\mathbb{F}_p$, then $A_1^{\perp}=A_2^{\perp}=0$ and 

\begin{align*}
A_3^{\perp}=(p-1)p^{2m-6}\frac{(p-2)p^{2m-6}+(p-1)^2p^{m-3}-p^2+p+1}{6}>0.
\end{align*}

{\bf Case 2.} If $4\mid m$ and $a\in\mathbb{F}_{p^2}\backslash\mathbb{F}_p$, then $A_1^{\perp}=A_2^{\perp}=0$ and 

\begin{align*}
A_3^{\perp}=\frac{1}{6}p^{m-4}(p-1)^2\big(p^{m-1}-p^{m-2}-p^2+3\big)>0.
\end{align*}

{\bf Case 3.}  If $m$ is even and $a\in\mathbb{F}_{p^{2}}\backslash\mathbb{F}_{p}$ with $\eta_{m}(a)=-1$, then $A_1^{\perp}=A_2^{\perp}=0$  and 

\begin{align*}
A_3^{\perp}=\frac{1}{6}p^{m-4}(p-1)^2\big(p^{m-1}-p^{m-2}-p+1\big)>0.
\end{align*}

{\bf Case 4.} If $m$ is even,  $s>2$ with $s\mid \frac{m}{2}$ and $a\in\mathbb{F}_{p^{s}}\backslash\mathbb{F}_{p^2}$, then $A_1^{\perp}=0$ and 

\begin{align*}
A_2^{\perp}=\frac{1}{2}p^{m-4}(p-1)^2>0.
\end{align*}$\hfill\Box$

Some examples for Theorems \ref{c2}-\ref{c4} are given as follows, respectively.

\begin{example}\label{e2}
	Let $p=3$, $m=8$ and $\mathbb{F}_{3^8}^{*}=\langle\beta \rangle$, where the minimal polynomial  over $\mathbb{F}_3$ of $\beta$ is $x^8+2x^5+x^4+2x^2+2x+2$ , for $a=\beta^{820}\in\mathbb{F}_{3^2}\backslash\mathbb{F}_{3}$, by $(\ref{C1})$, using MAGMA program, we get that $\mathcal{C}_{{D_a}}$ is a $[729,8,468]$ linear code with the weight enumerator $$1 +720z^{468}+2178z^{477}+728z^{486}+2196z^{495}+738z^{504}$$ and the complete weight enumerator
		\begin{align*}1&+728w_0^{243}w_1^{243}w_2^{243}+1098w_0^{234}w_1^{261}w_2^{234}+1098w_0^{234}w_1^{234}w_2^{261}\\
		&+1089w_0^{252}w_1^{225}w_2^{252}+1089w_0^{252}w_1^{252}w_2^{225}+720w_0^{261}w_1^{234}w_2^{234}+738w_0^{225}w_1^{252}w_2^{252},
		\end{align*}
	 which is accordant with Theorem $\ref{c2}$. 
\end{example}

\begin{example}\label{e3}
	Let $p=3$, $m=8$, $s=4$ and $\mathbb{F}_{3^8}^{*}=\langle\beta \rangle$, where the minimal polynomial of $\beta$  over $\mathbb{F}_3$ is $x^8+2x^5+x^4+2x^2+2x+2$, for $a=\beta^{82}\in\mathbb{F}_{3^4}\backslash\mathbb{F}_{3^2}$, by $(\ref{C1})$, using MAGMA program, we get that $\mathcal{C}_{{D_a}}$ is a $[729,8,459]$ linear code with the weight enumerator $$1 +240z^{459}+720z^{468}+1458z^{477}+1700z^{486}+1458z^{495}+738z^{504}+246z^{513}$$ and the complete weight enumerator
	 \begin{align*}1&+1700w_0^{243}w_1^{243}w_2^{243}+123w_0^{216}w_1^{243}w_2^{270}+123w_0^{216}w_1^{270}w_2^{243}\\
	 &+120w_0^{270}w_1^{243}w_2^{216}+120w_0^{270}w_1^{216}w_2^{243}+720w_0^{261}w_1^{234}w_2^{234}+738w_0^{225}w_1^{252}w_2^{252}\\
	 &+729w_0^{234}w_1^{234}w_2^{261}+729w_0^{234}w_1^{261}w_2^{234}+729
	 w_0^{252}w_1^{225}w_2^{252}+729w_0^{252}w_1^{252}w_2^{225},
	 \end{align*}
	which is accordant with Theorem $\ref{c3}$. 
\end{example}
\begin{example}\label{e4}
	Let $p=3$, $m=6$, and $\mathbb{F}_{3^6}^{*}=\langle\beta \rangle$, where the minimal polynomial of $\beta$  over $\mathbb{F}_3$ is $x^6+2x^4+x^2+2x+2$, for $a=\beta^{91}\in\mathbb{F}_{3^2}\backslash\mathbb{F}_{3}$, by $(\ref{C1})$, using MAGMA program, we get that $\mathcal{C}_{{D_a}}$ is a $[81,6,48]$ linear code with the weight enumerator $$1+90z^{48}+216z^{51}+80z^{54}+270z^{57}+72z^{60}$$ and the complete weight enumerator
	\begin{align*}1&+80w_0^{27}w_1^{27}w_2^{27}+135w_0^{24}w_1^{33}w_2^{24}+135w_0^{24}w_1^{24}w_2^{33}\\
	&+108w_0^{30}w_1^{21}w_2^{30}+108w_0^{30}w_1^{30}w_2^{21}+90w_0^{33}w_1^{24}w_2^{24}+72w_0^{21}w_1^{30}w_2^{30},
	\end{align*}
	 which is accordant with Theorem $\ref{c4}$. 
\end{example}
$\hfill\Box$
	\section{Proofs for Main Results}
	\subsection{Some auxiliary lemmas}
	
	Lemmas \ref{l41}-\ref{l42} are useful for calculating the length and the weights for codewords in $\mathcal{C}_{D_a}$.
	
	\begin{lemma}\label{l41}
			For any integer $m\ge 3$ and $a\in\mathbb{F}_{p^{m}}\backslash\mathbb{F}_p$, denote
		\begin{align*}
		D_{a}=\{ x\in\mathbb{F}_{p^m} ~\big|~\mathrm{Tr}(x)=1~\text{and}~\mathrm{Tr}(ax)=0\},
		\end{align*}
		then \begin{align*}
		\big|D_a\big|=p^{m-2}.
		\end{align*}
	\end{lemma}
	
	{\bf Proof.}~ By the orthogonal property for the additive character, we have
	\begin{align*}
	\big|D_{a}\big|
	=&\sum_{x\in\mathbb{F}_{p^m}}\Big(\frac{1}{p}\sum_{z_1\in\mathbb{F}_{p}}\zeta_p^{z_1(\mathrm{Tr}(x)-1)}\Big)\Big(\frac{1}{p}\sum_{z_2\in\mathbb{F}_{p}}\zeta_p^{z_2\mathrm{Tr}(ax)}\Big)\\
	=&\frac{1}{p^2}\sum_{z_2\in\mathbb{F}_{p}}\sum_{z_1\in\mathbb{F}_{p}}\zeta_p^{-z_1}\sum_{x\in\mathbb{F}_{p^m}}\zeta_p^{\mathrm{Tr}((z_1+z_2a)x)}\\
	=&p^{m-2}.
	\end{align*}$\hfill\Box$\\
	
	In the following lemma, we denote $\mathrm{T}_a(b)=\frac{\mathrm{Tr}(b^{-1})\mathrm{Tr}(a^2b^{-1})-\mathrm{Tr}(ab^{-1})^2}{\mathrm{Tr}(a^2b^{-1})}$ for convenience.
	\begin{lemma}\label{l42}
			For any integer $m\ge 3$, $a\in\mathbb{F}_{p^{m}}\backslash\mathbb{F}_p$, $\rho\in \mathbb{F}_{p}$ and $b\in \mathbb{F}_{p^m}^{*}$, denote
		\begin{align*}
		N_{a}(b,\rho)= \{x\in \mathbb{F}_{{p^{m}}}\big|\mathrm{Tr}(bx^2)=\rho,~\mathrm{Tr}(x)=1,~\mathrm{Tr}(ax)=0\}
		\end{align*}
		then 
		
		$({\bf\uppercase\expandafter{\romannumeral1}})$  for any odd $m$, 
		
		$(1)$ if $\mathrm{Tr}(a^2b^{-1})=\mathrm{Tr}(ab^{-1})=\mathrm{Tr}(b^{-1})=0$, or $\mathrm{Tr}(a^2b^{-1})\neq 0$ and $\mathrm{T}_a(b)=0$, then
		\begin{align*}
		|N_{a}(b,\rho)|=p^{m-3};
		\end{align*} 
		
		$(2)$ if $\mathrm{Tr}(a^2b^{-1})=\mathrm{Tr}(ab^{-1})=0$ and $\mathrm{Tr}(b^{-1})\neq 0$, then
		\begin{align*}
		|N_{a}(b,\rho)|=\begin{cases}
		p^{m-3}+\eta_m(b)\eta_{1}\big(-\mathrm{Tr}(b^{-1})\big)(p-1) \frac{G_mG_1}{p^2},\quad& \rho=(\mathrm{Tr}(b^{-1}))^{-1};\\
		p^{m-3}-\eta_m(b)\eta_{1}\big(-\mathrm{Tr}(b^{-1})\big)\frac{G_mG_1}{p^2},\quad& \rho\neq(\mathrm{Tr}(b^{-1}))^{-1};
		\end{cases}
		\end{align*} 	
		
		$(3)$ if $\mathrm{Tr}(a^2b^{-1})=0$ and $\mathrm{Tr}(ab^{-1})\neq 0$, then
		\begin{align*}
		|N_{a}(b,\rho)|=\begin{cases}
		p^{m-3},\quad &\rho=0;\\
		p^{m-3}-\eta_m(b)\frac{\eta_{1}(-\rho)G_mG_1}{p^2},\quad &\rho\neq0;
		\end{cases}
		\end{align*} 	
		
		$(4)$ if $\mathrm{Tr}(a^2b^{-1})\neq 0$ and $\mathrm{T}_a(b)\neq 0$, then
		\begin{align*}
		|N_{a}(b,\rho)|=\begin{cases}
		p^{m-3},\quad& \rho=(\mathrm{T}_a(b))^{-1};\\
		p^{m-3}+\eta_m(b)\eta_{1}\big(\mathrm{Tr}(a^2b^{-1})\big)\eta_{1}\big(\mathrm{T}_a(b)\rho-1\big)\frac{G_mG_1}{p^2},\quad& \rho\neq(\mathrm{T}_a(b))^{-1}.
		\end{cases}
		\end{align*}\\	
		
		$({\bf \uppercase\expandafter{\romannumeral2}})$ For any even $ m$, 
		
		$(1)$ if $\mathrm{Tr}(a^2b^{-1})=\mathrm{Tr}(ab^{-1})=\mathrm{Tr}(b^{-1})=0$, or $\mathrm{Tr}(a^2b^{-1})\neq 0$ and $\mathrm{T}_a(b)=0$, then
		\begin{align*}
		|N_{a}(b,\rho)|=p^{m-3};
		\end{align*} 
		
		$(2)$ if $\mathrm{Tr}(a^2b^{-1})=\mathrm{Tr}(ab^{-1})=0$ and $\mathrm{Tr}(b^{-1})\neq 0$, then
		\begin{align*}
		|N_{a}(b,\rho)|=\begin{cases}
		p^{m-3},\quad& \rho=(\mathrm{Tr}(b^{-1}))^{-1};\\
		p^{m-3}+\eta_m(b)\eta_{1}(1-\rho \mathrm{Tr}(b^{-1}))\frac{G_m}{p},\quad& \rho\neq(\mathrm{Tr}(b^{-1}))^{-1};
		\end{cases}
		\end{align*} 	
		
		$(3)$ if $\mathrm{Tr}(a^2b^{-1})=0$ and $\mathrm{Tr}(ab^{-1})\neq 0$, then
		\begin{align*}
		|N_{a}(b,\rho)|=\begin{cases}
		p^{m-3}+\eta_m(b)(p-1)\frac{G_m}{p^2},\quad &\rho=0;\\
		p^{m-3}-\eta_m(b)\frac{G_m}{p^2},\quad &\rho\neq 0;
		\end{cases}
		\end{align*} 	
		
		$(4)$ if $\mathrm{Tr}(a^2b^{-1})\neq 0$ and  $\mathrm{T}_a(b)\neq 0$, then
		\begin{align*}
		|N_{a}(b,\rho)|=\begin{cases}
		p^{m-3}+\eta_{1}(-1)\eta_m(b)\eta_{1}\big(\mathrm{Tr}(a^2b^{-1})\big)\eta_1\big(\mathrm{T}_a(b)\big)(p-1)\frac{G_m}{p^2},\quad& \rho=(\mathrm{T}_a(b))^{-1};\\
		p^{m-3}-\eta_{1}(-1)\eta_m(b)\eta_{1}\big(\mathrm{Tr}(a^2b^{-1})\big)\eta_1\big(\mathrm{T}_a(b)\big)\frac{G_m}{p^2},\quad& \rho\neq(\mathrm{T}_a(b))^{-1}.
		\end{cases}
		\end{align*} 	
		
	\end{lemma}

	{\bf Proof.}~ By the orthogonal property for the additive character, we have
	\begin{align*}
	\big|N_{a}(b,\rho)\big|
	=& \sum_{x\in\mathbb{F}_{p^m}}
	\Big(\frac{1}{p}\sum_{z_0\in\mathbb{F}_{p}}\zeta_p^{z_0(\mathrm{Tr}(bx^2)-\rho)}\Big)\Big(\frac{1}{p}\sum_{z_1\in\mathbb{F}_{p}}\zeta_p^{z_1(\mathrm{Tr}(x)-1)}\Big)\Big(\frac{1}{p}\sum_{z_2\in\mathbb{F}_{p}}\zeta_p^{z_2\mathrm{Tr}(ax)}\Big)\\
	=&\frac{|D_a|}{p}+\frac{1}{p^3}\sum_{x\in\mathbb{F}_{p^m}}
	\sum_{z_1\in\mathbb{F}_{p}}\zeta_p^{z_1(\mathrm{Tr}(x)-1)}
	\sum_{z_2\in\mathbb{F}_{p}}\zeta_p^{z_2\mathrm{Tr}(ax)}\sum\limits_{z_0\in\mathbb{F}_{p}^{*}}
	\zeta_p^{z_0(\mathrm{Tr}(bx^2)-\rho)}\\	
	=&p^{m-3}+\frac{1}{p^3}\sum\limits_{z_0\in\mathbb{F}_{p}^{*}}\zeta_p^{-\rho z_0}\sum_{z_1\in\mathbb{F}_{p}}\zeta_p^{-z_1}
	\sum_{z_2\in\mathbb{F}_{p}}\sum_{x\in\mathbb{F}_{p^m}}\zeta_p^{\mathrm{Tr}(z_0bx^2+(z_1+z_2a)x)}\\
	=&p^{m-3}+\eta_{m}(b)\frac{G_m}{p^3}\sum\limits_{z_0\in\mathbb{F}_{p}^{*}}\eta_{m}(z_0)\zeta_p^{-\rho z_0}\sum_{z_1\in\mathbb{F}_{p}}\zeta_p^{-z_1}
	\sum_{z_2\in\mathbb{F}_{p}}\zeta_p^{\mathrm{Tr}\big(-\frac{(z_1+z_2a)^2}{4z_0 b}\big)}\\
	=&p^{m-3}+\eta_{m}(b)\frac{G_m}{p^3}\sum\limits_{z_0\in\mathbb{F}_{p}^{*}}\eta_{m}(z_0)\zeta_p^{-\rho z_0}
	\sum_{z_1\in\mathbb{F}_{p}}\zeta_p^{-\frac{\mathrm{Tr}(b^{-1})}{4z_0}z_1^2-z_1}\sum_{z_2\in\mathbb{F}_{p}}\zeta_p^{-\frac{\mathrm{Tr}(a^2b^{-1})}{4z_0}z_2^2-\frac{\mathrm{Tr}(ab^{-1})z_1}{2z_0}z_2}\\
	=&p^{m-3}+\eta_{m}(b)\frac{G_m}{p^3}\Omega,
	\end{align*}
	where
	\begin{align*}
	\Omega=\sum\limits_{z_0\in\mathbb{F}_{p}^{*}}\eta_{m}(z_0)\zeta_p^{-\rho z_0}
	\sum_{z_1\in\mathbb{F}_{p}}\zeta_p^{-\frac{\mathrm{Tr}(b^{-1})}{4z_0}z_1^2-z_1}\sum_{z_2\in\mathbb{F}_{p}}\zeta_p^{-\frac{\mathrm{Tr}(a^2b^{-1})}{4z_0}z_2^2-\frac{\mathrm{Tr}(ab^{-1})z_1}{2z_0}z_2}.
	\end{align*}
	
	Now we can calculate $\Omega$ by the following two cases.
	
	${\bf Case 1.}$ For $\mathrm{Tr}(a^2b^{-1})=0$, 
	\begin{align*}
	\Omega=\sum\limits_{z_0\in\mathbb{F}_{p}^{*}}\eta_{m}(z_0)\zeta_p^{-\rho z_0}
	\sum_{z_1\in\mathbb{F}_{p}}\zeta_p^{-\frac{\mathrm{Tr}(b^{-1})}{4z_0}z_1^2-z_1}\sum_{z_2\in\mathbb{F}_{p}}\zeta_p^{
		-\frac{\mathrm{Tr}(ab^{-1})z_1}{2z_0}z_2}.
	\end{align*}
	If $\mathrm{Tr}(ab^{-1})=\mathrm{Tr}(b^{-1})=0$, then
	\begin{align*}
	\Omega=0.
	\end{align*}
	If $\mathrm{Tr}(ab^{-1})=0$ and $\mathrm{Tr}(b^{-1})\neq 0$, then
	\begin{align*}
	\Omega&=p\sum\limits_{z_0\in\mathbb{F}_{p}^{*}}\eta_{m}(z_0)\zeta_p^{-\rho z_0}\sum_{z_1\in\mathbb{F}_{p}}\zeta_p^{-\frac{\mathrm{Tr}(b^{-1})}{4z_0}z_1^2-z_1}\\
	&=\eta_{1}\Big(-\mathrm{Tr}(b^{-1})\Big)pG_1\sum\limits_{z_0\in\mathbb{F}_{p}^{*}}\eta_{m}(z_0)\eta_{1}(z_0)\zeta_p^{\big((\mathrm{Tr}(b^{-1}))^{-1}-\rho\big)  z_0}\\
	&=\begin{cases}
	\eta_{1}\big(-\mathrm{Tr}(b^{-1})\big)pG_1\sum\limits_{z_0\in\mathbb{F}_{p}^{*}}\zeta_p^{\big((\mathrm{Tr}(b^{-1}))^{-1}-\rho\big)  z_0},\quad&\text{~if~}m\text{~is~odd};\\
	\eta_{1}\big(-\mathrm{Tr}(b^{-1})\big)pG_1\sum\limits_{z_0\in\mathbb{F}_{p}^{*}}\eta_{1}(z_0)\zeta_p^{\big((\mathrm{Tr}(b^{-1}))^{-1}-\rho\big)  z_0},\quad&\text{~if~}m\text{~is~even};\\
	\end{cases}\\
	&=\begin{cases}
	\eta_{1}\big(-\mathrm{Tr}(b^{-1})\big)(p-1)pG_1,\quad&\text{~if~}m\text{~is~odd~}\text{and}~\rho=(\mathrm{Tr}(b^{-1}))^{-1};\\
	-\eta_{1}\big(-\mathrm{Tr}(b^{-1})\big)pG_1,\quad&\text{~if~}m\text{~is~odd~}\text{and}~\rho\neq(\mathrm{Tr}(b^{-1}))^{-1};\\
	0,\quad&\text{~if~}m\text{~is~even~}\text{and}~\rho=(\mathrm{Tr}(b^{-1}))^{-1};\\
	\eta_{1}(\rho \mathrm{Tr}(b^{-1})-1)pG_1^2,\quad&\text{~if~}m\text{~is~even~}\text{and}~\rho\neq(\mathrm{Tr}(b^{-1}))^{-1}.
	\end{cases}
	\end{align*}
	If $\mathrm{Tr}(ab^{-1})\neq0$, then
	\begin{align*}
	\Omega&=p\sum\limits_{z_0\in\mathbb{F}_{p}^{*}}\eta_{m}(z_0)\zeta_p^{-\rho z_0}\\
	&=
	\begin{cases}
	p\sum\limits_{z_0\in\mathbb{F}_{p}^{*}}\eta_{1}(z_0)\zeta_p^{-\rho z_0},\quad& \text{~if~}m\text{~is~odd};\\
	p\sum\limits_{z_0\in\mathbb{F}_{p}^{*}}\zeta_p^{-\rho z_0},\quad&\text{~if~}m\text{~is~even};\\
	\end{cases}\\
	&=
	\begin{cases}
	0,\quad& \text{~if~}m\text{~is~odd~}\text{and}~\rho=0;\\
	\eta_{1}(-\rho)pG_1,\quad&\text{~if~}m\text{~is~odd~}\text{and}~\rho\neq0;\\
	(p-1)p,\quad& \text{~if~}m\text{~is~even~}\text{and}~\rho=0;\\
	-p,\quad&\text{~if~}m\text{~is~even~}\text{and}~\rho\neq0.
	\end{cases}
	\end{align*}

	${\bf Case 2.}$ For $\mathrm{Tr}(a^2b^{-1})\neq 0$, by Lemma \ref{l23}, we have
	\begin{align*}
	\Omega
	=&\sum\limits_{z_0\in\mathbb{F}_{p}^{*}}\eta_{m}(z_0)\zeta_p^{-\rho z_0}
	\sum_{z_1\in\mathbb{F}_{p}}\zeta_p^{-\frac{\mathrm{Tr}(b^{-1})}{4z_0}z_1^2-z_1}\sum_{z_2\in\mathbb{F}_{p}}\zeta_p^{-\frac{\mathrm{Tr}(a^2b^{-1})}{4z_0}z_2^2-\frac{\mathrm{Tr}(ab^{-1})z_1}{2z_0}z_2}\\
	=&\sum\limits_{z_0\in\mathbb{F}_{p}^{*}}\eta_{m}(z_0)\zeta_p^{-\rho z_0}
	\sum_{z_1\in\mathbb{F}_{p}}\zeta_p^{-\frac{\mathrm{Tr}(b^{-1})}{4z_0}z_1^2-z_1}\bigg(G_1\eta_{1}\Big( -\frac{\mathrm{Tr}(a^2b^{-1})}{4z_0}\Big)\zeta_p^{\frac{\mathrm{Tr}(ab^{-1})^2}{4\mathrm{Tr}(a^2b^{-1})z_0}z_1^2}\bigg)\\
	=&\eta_{1}\big(-\mathrm{Tr}(a^2b^{-1})\big)G_1\sum\limits_{z_0\in\mathbb{F}_{p}^{*}}\eta_{m}(z_0)\eta_{1}(z_0)\zeta_p^{-\rho z_0}
	\sum_{z_1\in\mathbb{F}_{p}}\zeta_p^{-\frac{\mathrm{T}_a(b)}{4z_0}z_1^2-z_1}.	
	\end{align*}
	If $\mathrm{T}_a(b)= 0$, then
	\begin{align*}
	\Omega=0.
	\end{align*}
	If $\mathrm{T}_a(b)\neq0$, then
	\begin{align*}
	\Omega
	=&\eta_{1}\big(\mathrm{Tr}(a^2b^{-1})\big)\eta_1\big(\mathrm{T}_a(b)\big)G_1^2\sum\limits_{z_0\in\mathbb{F}_{p}^{*}}\eta_{m}(z_0)\zeta_p^{\big((\mathrm{T}_a(b))^{-1}-\rho\big) z_0}\\
	=&\begin{cases}
	\eta_{1}\big(\mathrm{Tr}(a^2b^{-1})\big)\eta_1\big(\mathrm{T}_a(b)\big)G_1^2\sum\limits_{z_0\in\mathbb{F}_{p}^{*}}\eta_{1}(z_0)\zeta_p^{\big((\mathrm{T}_a(b))^{-1}-\rho\big) z_0},\quad &\text{~if~}m\text{~is~odd};\\
	\eta_{1}\big(\mathrm{Tr}(a^2b^{-1})\big)\eta_1\big(\mathrm{T}_a(b)\big)G_1^2\sum\limits_{z_0\in\mathbb{F}_{p}^{*}}\zeta_p^{\big((\mathrm{T}_a(b))^{-1}-\rho\big) z_0},\quad &\text{~if~}m\text{~is~even};\\		
	\end{cases}\\
	=&\begin{cases}
	0,\quad &\text{~if~}m\text{~is~odd~}\text{and}~\rho=(\mathrm{T}_a(b))^{-1};\\
	\eta_{1}\big(\mathrm{Tr}(a^2b^{-1})\big)\eta_{1}\big(1-\mathrm{T}_a(b)\rho\big)G_1^3,\quad &\text{~if~}m\text{~is~odd~}\text{and}~\rho\neq(\mathrm{T}_a(b))^{-1};\\
	\eta_{1}\big(\mathrm{Tr}(a^2b^{-1})\big)\eta_1\big(\mathrm{T}_a(b)\big)(p-1)G_1^2,\quad &\text{~if~}m\text{~is~even~}\text{and}~\rho=(\mathrm{T}_a(b))^{-1};\\
	-\eta_{1}\big(\mathrm{Tr}(a^2b^{-1})\big)\eta_1\big(\mathrm{T}_a(b)\big)G_1^2,\quad &\text{~if~}m\text{~is~even~}\text{and}~\rho\neq(\mathrm{T}_a(b))^{-1}.
	\end{cases}\\	
	\end{align*}
	
	Now, by the above discussions  and using  $p=\eta_{1}(-1)G_1^2$, Lemma \ref{l42} is immediate. $\hfill\Box$\\
	
	In order to calculate the weight enumerator for $\mathcal{C}_{D_a}$, it is necessary to count the number of elements for the following four sets,
		{\footnotesize \begin{align*}
			&{M}_{1,a}=\Big\{ b\in\mathbb{F}_{p^m}^{*}\big|\mathrm{Tr}(a^2b^{-1})=\mathrm{Tr}(ab^{-1})=\mathrm{Tr}(b^{-1})=0, \text{or~}\mathrm{Tr}(a^2b^{-1})\neq 0\\
		&\qquad\qquad\qquad\qquad\qquad\qquad\qquad\qquad~~ \text{and~} \mathrm{Tr}(a^2b^{-1})\mathrm{Tr}(b^{-1})-\mathrm{Tr}(ab^{-1})^2=0 \Big \},\\
			&{M}_{2,a}(\epsilon,\gamma)=\left\{ b\in\mathbb{F}_{p^m}^{*}\big|\eta_{m}(b)=\epsilon,\mathrm{Tr}(a^2b^{-1})=\mathrm{Tr}(ab^{-1})=0\text{~and~}\mathrm{Tr}(b^{-1})=\gamma \right\},\\
			&{M}_{3,a}(\epsilon)=\left\{ b\in\mathbb{F}_{p^m}^{*}\big|\eta_{m}(b)=\epsilon,\mathrm{Tr}(a^2b^{-1})=0\text{~and~}\mathrm{Tr}(ab^{-1})\neq 0 \right\},\\
			&{M}_{4,a}(\epsilon,\gamma)\!=\!\left\{b\in\mathbb{F}_{p^m}^{*}\big|\eta_{m}(b)\eta_{1}\big(\mathrm{Tr}(a^2b^{-1})\big)=\epsilon~\text{and}~\frac{\mathrm{Tr}(a^2b^{-1})\mathrm{Tr}(b^{-1})-\mathrm{Tr}(ab^{-1})^2}{\mathrm{Tr}(a^2b^{-1})}=\gamma\right\}.
		\end{align*}}
	
	  Lemmas \ref{l43}-\ref{l46} are given to calculate  $|{M}_{1,a}|$, $|{M}_{2,a}(\epsilon,\gamma)|$, $|{M}_{3,a}(\epsilon)|$ and $|{M}_{4,a}(\epsilon,\gamma)|$.

	\begin{lemma}\label{l43}
			For any integer $m\ge 3$ and $a\in\mathbb{F}_{p^m}\backslash\mathbb{F}_p$, denote 
		\begin{align}
		M(a)=\{(z_0,z_1,z_2)\in\mathbb{F}_p^{3}~|~z_2a^2+z_1a+z_0\neq 0\},
		\end{align}
		then the following assertions hold.\\
		
		$(1)$ If  $m$ is odd, or $m$ is even and $a\in\mathbb{F}_{p^m}\backslash\mathbb{F}_{p^2}$ then 
		\begin{align*}
		M(a)=(\mathbb{F}_p^3)^{*}.
		\end{align*}
		
		$(2)$ If $m$ is even and $  a\in \mathbb{F}_{p^2}\backslash \mathbb{F}_p\text{~with~}a^2+b_1a+b_0=0~(b_1\in\mathbb{F}_p,b_0\in\mathbb{F}_p^{*})$, then
		\begin{align*}
		M(a)=\mathbb{F}_p^3\backslash\{(z_2b_0,z_2b_1,z_2)|z_2\in\mathbb{F}_p\}.
		\end{align*}
	\end{lemma}
	
	{\bf Proof.} If $m$ is odd, or $m$ is even and $a\in\mathbb{F}_{p^m}\backslash\mathbb{F}_{p^2}$,  then it is easy to see that 
	\begin{align*}
		z_2a^2+z_1a+z_0=0~\big((z_0,z_1,z_2)\in\mathbb{F}_p^{3}\big)\text{~if and only if~}z_0=z_1=z_2=0.
	\end{align*}
	 If $m$ is even and $a\in \mathbb{F}_{p^2}\backslash \mathbb{F}_p\text{~with~}a^2+b_1a+b_0=0~(b_1\in\mathbb{F}_p,b_0\in\mathbb{F}_p^{*})$, then we have
		\begin{align}\label{l431}
		z_2a^2+z_1a+z_0=0~\big((z_0,z_1,z_2)\in\mathbb{F}_p^{3}\big)\text{~if and only if~}(z_1-z_2b_1)a+(z_0-z_2b_0)=0,
		\end{align}
	and by $a\notin \mathbb{F}_p$, $(\ref{l431})$ is equivalent to both	$z_1=z_2b_1\text{~and~}z_0=z_2b_0$.
 $\hfill\Box$
	
		\begin{lemma}\label{l44}		For any integer $m\ge 3$, $a\in\mathbb{F}_{p^{m}}\backslash\mathbb{F}_p$ and $(\gamma_{0},\gamma_{1},\gamma_{2})\in\mathbb{F}_{p}^3$, denote 
			\begin{align*}
			T_a(\gamma_{0},\gamma_{1},\gamma_{2})=\bigg\{b\in\mathbb{F}_{p^m}^{*}~|~\mathrm{Tr}\big(b^{-1}\big)=\gamma_{0},~\mathrm{Tr}\big(ab^{-1}\big)=\gamma_{1}~\text{and}~\mathrm{Tr}({a^2}b^{-1})=\gamma_{2}\bigg\},
			\end{align*}
			then the following two assertions hold.\\
			
			$(1)$~If $m$ is odd , or $m$ is even~and~ $a\in\mathbb{F}_{p^m}\backslash\mathbb{F}_{p^2}$, then
			\begin{align*}
			\big|T_a(\gamma_{0},\gamma_{1},\gamma_{2})\big|=\begin{cases}
			p^{m-3}-1,\quad&\text{~if~}\gamma_{0}=\gamma_{1}=\gamma_{2}=0;\\
			p^{m-3},\quad& \text{~if~} (\gamma_{0},\gamma_{1},\gamma_{2})\in(\mathbb{F}_p^3)^{*}.
			\end{cases}
			\end{align*}
			
			$(2)$~If $m$ is even and $a\in\mathbb{F}_{p^2}\backslash\mathbb{F}_p$ with $a^2+b_1a+b_0=0$ $(b_1\in\mathbb{F}_p,b_0\in\mathbb{F}_p^{*})$, then
			\begin{align*}
			\big|T_a(\gamma_{0},\gamma_{1},\gamma_{2})\big|=\begin{cases}
			p^{m-2}-1,\quad&\text{~if~}\gamma_{0}=\gamma_{1}=\gamma_{2}=0;\\
			p^{m-2},\quad&\text{~if~}(\gamma_{0},\gamma_{1},\gamma_{2})\in(\mathbb{F}_p^3)^*\text{~and~}\gamma_{2}+b_1\gamma_{1}+b_0\gamma_{0}=0;\\
			0,\quad&\text{~if~}(\gamma_{0},\gamma_{1},\gamma_{2})\in(\mathbb{F}_p^3)^*\text{~and~}\gamma_{2}+b_1\gamma_{1}+b_0\gamma_{0}\neq 0.
			\end{cases}
			\end{align*}
		\end{lemma}	
		
	{\bf Proof.}~By the orthogonal property for the additive character,   we have \begin{align*}
	&\big|T_a(\gamma_{0},\gamma_{1},\gamma_{2})\big|\\
	=&\sum_{x\in\mathbb{F}_{p^m}^{*}}\big(\frac{1}{p}\sum_{z_1\in\mathbb{F}_{p}}\zeta_p^{z_1(\mathrm{Tr}(x^{-1})-\gamma_0)}\big)\big(\frac{1}{p}\sum_{z_1\in\mathbb{F}_{p}}\zeta_p^{z_1(\mathrm{Tr}(ax^{-1})-\gamma_{1})}\big)\big(\frac{1}{p}\sum_{z_2\in\mathbb{F}_{p}}\zeta_p^{z_2(\mathrm{Tr}(a^2x^{-1})-\gamma_{2})}\big)\\
	=&\sum_{x\in\mathbb{F}_{p^m}^{*}}\big(\frac{1}{p}\sum_{z_1\in\mathbb{F}_{p}}\zeta_p^{z_1(\mathrm{Tr}(x)-\gamma_0)}\big)\big(\frac{1}{p}\sum_{z_1\in\mathbb{F}_{p}}\zeta_p^{z_1(\mathrm{Tr}(ax)-\gamma_{1})}\big)\big(\frac{1}{p}\sum_{z_2\in\mathbb{F}_{p}}\zeta_p^{z_2(\mathrm{Tr}(a^2x)-\gamma_{2})}\big)\\
	=&\frac{1}{p^3}\Big(\sum_{z_2\in\mathbb{F}_{p}}\sum_{z_1\in\mathbb{F}_{p}}\sum_{z_0\in\mathbb{F}_{p}}\zeta_p^{-z_2\gamma_2-z_1\gamma_1-z_0\gamma_0}\sum_{x\in\mathbb{F}_{p^m}}\zeta_p^{\mathrm{Tr}((z_2a^2+z_1a+z_0)x)}-\sum_{z_2\in\mathbb{F}_{p}}\sum_{z_1\in\mathbb{F}_{p}}\sum_{z_0\in\mathbb{F}_{p}}\zeta_p^{-z_2\gamma_2-z_1\gamma_1-z_0\gamma_0}\Big)\\
	=&p^{m-3}\sum_{\substack{(z_0,z_1,z_2)\in (\mathbb{F}_p)^3\\z_2a^2+z_1a+z_0=0}}\zeta_p^{-z_2\gamma_2-z_1\gamma_1-z_0\gamma_0}-\frac{1}{p^3}\sum_{z_2\in\mathbb{F}_{p}}\sum_{z_1\in\mathbb{F}_{p}}\sum_{z_0\in\mathbb{F}_{p}}\zeta_p^{-z_2\gamma_2-z_1\gamma_1-z_0\gamma_0}\\
	=&\begin{cases}
	p^{m-3}(p^3-|M(a)|)-1,	\quad &\text{~if~} \gamma_{0}=\gamma_{1}=\gamma_{2}=0;\\
	-p^{m-3}\sum\limits_{(z_0,z_1,z_2)\in M(a)}\zeta_p^{-z_2\gamma_2-z_1\gamma_1-z_0\gamma_0},	\quad & \text{~if~} (\gamma_{0},\gamma_{1},\gamma_{2})\in(\mathbb{F}_p^3)^{*}.
	\end{cases}
	\end{align*}
	Now by Lemma \ref{l43}, if $m$ is odd, or $m$ is even and $a\in\mathbb{F}_{p^m}\backslash\mathbb{F}_{p^2}$, we have $M(a)=(\mathbb{F}_p^3)^{*}$, and then
	\begin{align*}
	\big|T_a(\gamma_{0},\gamma_{1},\gamma_{2})\big|=\begin{cases}
	p^{m-3}-1,\quad&\text{~if~} \gamma_{0}=\gamma_{1}=\gamma_{2}=0;\\
	p^{m-3},\quad&\text{~if~} (\gamma_{0},\gamma_{1},\gamma_{2})\in(\mathbb{F}_p^3)^{*}.
	\end{cases}
	\end{align*} 
	And for even $ m$ and $a\in\mathbb{F}_{p^2}\backslash\mathbb{F}_p$ with $a^2+b_1a+b_0=0$ $(b_1\in\mathbb{F}_p,b_0\in\mathbb{F}_p^{*})$, we have $M(a)=\mathbb{F}_p^3\backslash\{(z_2b_0,z_2b_1,z_2)|z_2\in\mathbb{F}_p\}$, and then
	\begin{align*}
	\big|T_a(\gamma_{0},\gamma_{1},\gamma_{2})\big|
	=&\begin{cases}
	p^{m-2}-1,\quad&\text{~if~} \gamma_{0}=\gamma_{1}=\gamma_{2}=0;\\
	p^{m-3}\sum\limits_{z_2\in \mathbb{F}_p}\zeta_p^{-(\gamma_2+b_1\gamma_1+b_0\gamma_0)z_2},\quad&	\text{~if~} (\gamma_{0},\gamma_{1},\gamma_{2})\in(\mathbb{F}_p^3)^{*};
	\end{cases}\\
	=&\begin{cases}
	p^{m-2}-1,\quad&\text{~if~} \gamma_{0}=\gamma_{1}=\gamma_{2}=0;\\
	p^{m-2},\quad&\text{~if~} (\gamma_{0},\gamma_{1},\gamma_{2})\in(\mathbb{F}_p^3)^*\text{~and~}\gamma_{2}+b_1\gamma_{1}+b_0\gamma_{0}=0;\\
	0,\quad&\text{~if~} (\gamma_{0},\gamma_{1},\gamma_{2})\in(\mathbb{F}_p^3)^*\text{~and~}\gamma_{2}+b_1\gamma_{1}+b_0\gamma_{0}\neq 0.
	\end{cases}
	\end{align*} $\hfill\Box$\\
		\begin{lemma}\label{l45}
			For any integer $m\ge 3$, $a\in\mathbb{F}_{p^{m}}\backslash\mathbb{F}_p$ and $(\gamma_{0},\gamma_{1},\gamma_{2})\in\mathbb{F}_{p}^3$, denote 
		\begin{align*}
		L_a(\epsilon,\gamma_{0},\gamma_{1},\gamma_{2})=\bigg\{b\in\mathbb{F}_{p^m}^{*}~|~\eta_{m}(b)=\epsilon,~~\mathrm{Tr}\big(b^{-1}\big)=\gamma_{0},~\mathrm{Tr}\big(ab^{-1}\big)=\gamma_{1}~\text{and}~\mathrm{Tr}({a^2}b^{-1})=\gamma_{2}\bigg\},
		\end{align*}
		then \begin{align*}|L_a(\epsilon,\gamma_{0},\gamma_{1},\gamma_{2})|=\frac{|T_a(\gamma_{0},\gamma_{1},\gamma_{2})|}{2}+\epsilon\cdot\frac{ G_m}{2p^3}\sum\limits_{(z_0,z_1,z_2)\in M(a)}\zeta_p^{-\gamma_{2}z_2-\gamma_{1}z_1-\gamma_{0}z_0}\eta_{m}(z_2a^2+z_1a+z_0).\end{align*}	
		\end{lemma}
	
		{\bf Proof}.~By the orthogonal property for the additive character, we have
		\begin{align*}
		&|L_a(\epsilon,\gamma_{0},\gamma_{1},\gamma_{2})|\\=&
		\sum_{x\in\mathbb{F}_{p^m}^{*}}\frac{1+\epsilon\cdot\eta_{m}(x^{-1})}{2}\Big(\frac{1}{p}\sum_{z_2\in\mathbb{F}_p}\zeta_p^{z_2(\mathrm{Tr}(a^2x^{-1})-\gamma_{2})}\Big)\Big(\frac{1}{p}\sum_{z_1\in\mathbb{F}_p}\zeta_p^{z_1(\mathrm{Tr}(ax^{-1})-\gamma_{1})}\Big)\Big(\frac{1}{p}\sum_{z_0\in\mathbb{F}_p}\zeta_p^{z_0(\mathrm{Tr}(x^{-1})-\gamma_{0})}\Big)\\
		=&
		\sum_{x\in\mathbb{F}_{p^m}^{*}}\frac{1+\epsilon\cdot\eta_{m}(x)}{2}\Big(\frac{1}{p}\sum_{z_2\in\mathbb{F}_p}\zeta_p^{z_2(\mathrm{Tr}(a^2x)-\gamma_{2})}\Big)\Big(\frac{1}{p}\sum_{z_1\in\mathbb{F}_p}\zeta_p^{z_1(\mathrm{Tr}(ax)-\gamma_{1})}\Big)\Big(\frac{1}{p}\sum_{z_0\in\mathbb{F}_p}\zeta_p^{z_0(\mathrm{Tr}(x)-\gamma_{0})}\Big)\\
		=& \frac{|T_a(\gamma_{0},\gamma_{1},\gamma_{2})|}{2}+\frac{\epsilon}{2p^3}\sum_{z_2\in\mathbb{F}_p}\zeta_p^{-\gamma_{2}z_2}\sum_{z_1\in\mathbb{F}_p}\zeta_p^{-\gamma_{1}z_1}\sum_{z_0\in\mathbb{F}_p}\zeta_p^{-\gamma_{0}z_0}\sum_{x\in\mathbb{F}_{p^m}^{*}}\eta_{m}(x)\zeta_p^{\mathrm{Tr}((z_2a^2+z_1a+z_0)x)}\\
		=& \frac{|T_a(\gamma_{0},\gamma_{1},\gamma_{2})|}{2}+\epsilon\cdot\frac{G_m}{2p^3}\sum_{(z_0,z_1,z_2)\in M(a)}\zeta_p^{-\gamma_{2}z_2-\gamma_{1}z_1-\gamma_{0}z_0}\eta_{m}(z_2a^2+z_1a+z_0).
		\end{align*} $\hfill\Box$
	
		\begin{lemma}
			For any integer $m\ge3$, $a\in\mathbb{F}_{p^m}\backslash\mathbb{F}_p$ and $\gamma\in\mathbb{F}_p^{*}$, denote
		\begin{align*}E_{1,a}(\gamma)=\sum_{\gamma_{1}\in\mathbb{F}_p}\sum\limits_{\gamma_{2}\in\mathbb{F}_{p}^{*}}\eta_{1}(\gamma_{2})\sum\limits_{(z_0,z_1,z_2)\in (\mathbb{F}_p^3)^{*}}\zeta_p^{-\gamma_{2}z_2-\gamma_{1}z_1-(\gamma_{1}^2\gamma_{2}^{-1}+\gamma)z_0}\eta_{m}(z_2a^2+z_1a+z_0),
		\end{align*}
		then \begin{align*}E_{1,a}(\gamma)=\begin{cases}	
		\eta_{1}(-\gamma)\big(p^2-I_{2}(a)\big)p,	\quad &\text{~if~}m\text{~is even};\\
		\eta_1(-1)\big(I_{2}(a)-p\big)G_1,\quad &\text{~if~}m\text{~is odd}.
		\end{cases}
		\end{align*}
	\end{lemma}

	{\bf Proof.}~By Lemmas \ref{l21}-\ref{l23}, we have{
		\begin{align}\label{l441}\begin{aligned}
		&E_{1,a}(\gamma)\\
		=&\sum_{\gamma_{1}\in\mathbb{F}_p}\sum\limits_{\gamma_{2}\in\mathbb{F}_{p}^{*}}\eta_{1}(\gamma_{2})\sum\limits_{(z_0,z_1,z_2)\in(\mathbb{F}_p^3)^{*}}\zeta_p^{-\gamma_{2}z_2-\gamma_{1}z_1-(\gamma_{1}^2\gamma_{2}^{-1}+\gamma)z_0}\eta_{m}(z_2a^2+z_1a+z_0)\\
		=&\sum_{\gamma_{1}\in\mathbb{F}_p}\sum\limits_{\gamma_{2}\in\mathbb{F}_{p}^{*}}\eta_{1}(\gamma_{2})\sum\limits_{(z_1,z_2)\in(\mathbb{F}_p^2)^{*}}\zeta_p^{-\gamma_{2}z_2-\gamma_{1}z_1}\eta_{m}(z_2a^2+z_1a)\\&+\sum\limits_{z_0\in\mathbb{F}_p^{*}}\sum\limits_{z_1\in\mathbb{F}_p}\sum\limits_{z_2\in\mathbb{F}_p}\sum\limits_{\gamma_{2}\in\mathbb{F}_{p}^{*}}\eta_{1}(\gamma_{2})\sum_{\gamma_{1}\in\mathbb{F}_p}\zeta_p^{-z_0\gamma_{2}^{-1}\gamma_{1}^2-z_1\gamma_{1}-z_2\gamma_{2}-\gamma z_0}\eta_{m}(z_2a^2+z_1a+z_0)\\
		=&p\sum\limits_{z_2\in\mathbb{F}_p^*}\sum\limits_{\gamma_{2}\in\mathbb{F}_{p}^{*}}\eta_{1}(\gamma_{2})\zeta_p^{-\gamma_{2}z_2}\eta_{m}(z_2)\\&+G_1\sum\limits_{z_0\in\mathbb{F}_p^{*}}\sum\limits_{z_1\in\mathbb{F}_p}\sum\limits_{z_2\in\mathbb{F}_p}\sum\limits_{\gamma_{2}\in\mathbb{F}_{p}^{*}}\zeta_p^{\frac{z_1^2-4z_0z_2}{4z_0}\gamma_{2}-\gamma z_0}\eta_{1}(-z_0)\eta_{m}(z_2a^2+z_1a+z_0)\\
		=&pG_1\sum\limits_{z_2\in\mathbb{F}_p^*}\eta_{1}(-z_2)\eta_{m}(z_2)-G_1\sum\limits_{z_0\in\mathbb{F}_p^{*}}\sum\limits_{z_1\in\mathbb{F}_p}\sum\limits_{z_2\in\mathbb{F}_p}\zeta_p^{-\gamma z_0}\eta_{1}(-z_0)\eta_{m}(z_2a^2+z_1a+z_0)\\		&+G_1\sum\limits_{z_0\in\mathbb{F}_p^{*}}\sum\limits_{z_1\in\mathbb{F}_p}\sum\limits_{z_2\in\mathbb{F}_p}\sum\limits_{\gamma_{2}\in\mathbb{F}_{p}}\zeta_p^{\frac{z_1^2-4z_0z_2}{4z_0}\gamma_{2}}\zeta_p^{-\gamma z_0}\eta_{1}(-z_0)\eta_{m}(z_2a^2+z_1a+z_0)\\	
		=&pG_1\sum\limits_{z_2\in\mathbb{F}_p^*}\eta_{1}(-z_2)\eta_{m}(z_2)-G_1\sum\limits_{z_0\in\mathbb{F}_p^{*}}\zeta_p^{-\gamma z_0}\eta_{1}(-z_0)\eta_{m}(z_0)\sum\limits_{z_1\in\mathbb{F}_p}\sum\limits_{z_2\in\mathbb{F}_p}\eta_{m}(z_2a^2+z_1a+1)\\&+pG_1\sum\limits_{z_0\in\mathbb{F}_p^{*}}\zeta_p^{-\gamma z_0}\eta_{1}(-z_0)\eta_{m}(z_0)+pG_1\sum\limits_{z_1\in\mathbb{F}_p^{*}}\sum\limits_{z_2\in\mathbb{F}_p^*}\zeta_p^{-\gamma \frac{z_1^2}{4z_2}}\eta_{1}(-\frac{z_1^2}{4z_2})\eta_{m}(z_2a^2+z_1a+\frac{z_1^2}{4z_2})\\
		=&pG_1\Big(\sum\limits_{z_2\in\mathbb{F}_p^*}\eta_{1}(-z_2)\eta_{m}(z_2)+\sum\limits_{z_0\in\mathbb{F}_p^{*}}\zeta_p^{-\gamma z_0}\eta_{1}(-z_0)\eta_{m}(z_0)+\sum\limits_{z_1\in\mathbb{F}_p^{*}}\sum\limits_{z_2\in\mathbb{F}_p^*}\zeta_p^{-\gamma \frac{z_1^2}{4z_2}}\eta_{1}(-z_2)\eta_{m}(z_2)\Big)\\
		&-I_{2}(a)G_1\sum\limits_{z_0\in\mathbb{F}_p^{*}}\zeta_p^{-\gamma z_0}\eta_{1}(-z_0)\eta_{m}(z_0)\\	
		=&\begin{cases}
		pG_1\Big(\sum\limits_{z_0\in\mathbb{F}_p^{*}}\zeta_p^{-\gamma z_0}\eta_{1}(-z_0)+\sum\limits_{z_1\in\mathbb{F}_p^{*}}\!\sum\limits_{z_2\in\mathbb{F}_p^*}\!\zeta_p^{-\gamma \frac{z_1^2}{4z_2}}\eta_{1}(-z_2)\!\Big)-	I_{2}(a)G_1\!\!\sum\limits_{z_0\in\mathbb{F}_p^{*}}\zeta_p^{-\gamma z_0}\eta_{1}(-z_0),\!\! \!\!&\quad \text{~if~}m\text{~is even};\\
		\eta_{1}(-1)pG_1\Big(p-1+\!\sum\limits_{z_0\in\mathbb{F}_p^{*}}\!\zeta_p^{-\gamma z_0}+\!\sum\limits_{z_1\in\mathbb{F}_p^{*}}\sum\limits_{z_2\in\mathbb{F}_p^*}\zeta_p^{-\gamma \frac{z_1^2}{4z_2}}\Big)
		-\eta_1(-1)I_{2}(a)G_1\!\!\sum\limits_{z_0\in\mathbb{F}_p^{*}}\zeta_p^{-\gamma z_0},\!\!\!\! &\quad\text{~if~}m\text{~is odd};
		\end{cases}\\
		=&\begin{cases}
		\eta_{1}(\gamma)G_1^2\big(p^2-I_{2}(a)\big),&\quad \text{~if~}m\text{~is even};\\
		\eta_1(-1)G_1\big(I_{2}(a)-p\big),&\quad \text{~if~}m\text{~is odd}.
		\end{cases}
		\end{aligned}
		\end{align}}
	
		Now by $(\ref{l441})$ and $p=\eta_{1}(-1)G_1^2$, we complete the proof.
	$\hfill\Box$
	\begin{lemma}\label{l46}
		For any even $m$ and $a\in\mathbb{F}_{p^2}\backslash\mathbb{F}_p$ with $a^2+b_1a+b_0$ $(b_1\in\mathbb{F}_p,b_0\in\mathbb{F}_p^*)$, denote{
		\begin{align*}E_{2,a}(\gamma)=
		\sum_{\gamma_{1}\in\mathbb{F}_p}\sum\limits_{\gamma_{2}\in\mathbb{F}_{p}^{*}}\eta_{1}(\gamma_{2})\sum\limits_{\substack{(z_0,z_1,z_2)\in \mathbb{F}_p^{3}\\(z_1,z_0)=(z_2b_1,z_2b_0)}}\zeta_p^{-\gamma_{2}z_2-\gamma_{1}z_1-(\gamma_{1}^2\gamma_{2}^{-1}+\gamma)z_0}\eta_{m}(z_2a^2+z_1a+z_0),
		\end{align*}}
		then  \begin{align*}E_{2,a}(\gamma)=0.
		\end{align*}
	\end{lemma}

	{\bf Proof.}~For even $ m$ and $a\in\mathbb{F}_{p^2}\backslash\mathbb{F}_p$ with $a^2+b_1a+b_0=0$, we have
	{\small
		\begin{align*}
		E_{2,a}(\gamma)	=&\sum\limits_{z_2\in\mathbb{F}_p^{*}}\sum_{\gamma_{1}\in\mathbb{F}_p}\sum\limits_{\gamma_{2}\in\mathbb{F}_{p}^{*}}\eta_{1}(\gamma_{2})\zeta_p^{-z_2b_0\gamma_{2}^{-1}\gamma_{1}^2-z_2b_1\gamma_{1}-z_2\gamma_{2}+\gamma b_0 z_2}\eta_{m}\big(z_2(a^2+b_1a+b_0)\big)=0.
		\end{align*}
	}$\hfill\Box$
		
	Now, we give the values of $|{M}_{1,a}|$, $|{M}_{2,a}(\epsilon,\gamma)|$, $|{M}_{3,a}(\epsilon)|$ and $|{M}_{4,a}(\epsilon,\gamma)|$ in Lemmas \ref{l47}-\ref{l410}, respectively.
	
	\begin{lemma}\label{l47}
			For any integer $m\ge 3$ and $a\in\mathbb{F}_{p^{m}}\backslash\mathbb{F}_p$, we have
		\begin{align*}
			|M_{1,a}|=\begin{cases}
			p^{m-1}-p^{m-2}+p^{m-3}-1,\quad& \text{~if~} m \text{~is~odd} , \text{~or~} m \text{~is~even} \text{~and~} a\in\mathbb{F}_{p^m}\backslash\mathbb{F}_{p^2};\\
			p^{m-2}-1,\quad&\text{~if~} m \text{~is~even} \text{~and~}a\in\mathbb{F}_{p^2}\backslash\mathbb{F}_p.
			\end{cases}
		\end{align*}
		
	\end{lemma}

	{\bf Proof.}~By calculating directly, we have
	\begin{align*}
	M_{1,a}&= T_a(0,0,0)\bigcup\Bigg(\bigcup_{\gamma_{2}\in\mathbb{F}_p^{*}}\bigcup_{\gamma_{1}\in\mathbb{F}_p}T_a(\gamma_{1}^2\gamma_{2}^{-1},\gamma_{1},\gamma_{2})\Bigg).
	\end{align*}
	And now, by Lemma $\ref{l44}$, $|M_{1,a}|$ is given by the following two cases.
	
	$(1)$~If $m$ is odd, or $m$ is even and $a\in\mathbb{F}_{p^m}\backslash\mathbb{F}_{p^2}$, 
	\begin{align*}
	|M_{1,a}|=|T_a(0,0,0)|+\sum_{\gamma_{2}\in\mathbb{F}_p^{*}}\sum_{\gamma_{1}\in\mathbb{F}_p}|T_a(\gamma_{1}^2\gamma_{2}^{-1},\gamma_{1},\gamma_{2})|
	=p^{m-3}-1+p^{m-1}-p^{m-2}.
	\end{align*}
	
	$(2)$~If $m$ is even and $a\in\mathbb{F}_{p^2}\backslash\mathbb{F}_p$, then there exist some $b_1\in\mathbb{F}_p$ and $b_0\in\mathbb{F}_p^{*}$ such that $a^2+b_1a+b_0=0$, and then by $a\notin\mathbb{F}_p$, we know that
	\begin{align}\label{l471}
		x^2+b_1x+b_0\neq 0\quad \text{for all}~x\in\mathbb{F}_p. 
	\end{align}
	Now by $(\ref{l471})$ and Lemma $\ref{l44}$, we have
	\begin{align*}
	|M_{1,a}|=&|T_a(0,0,0)|+\sum_{\gamma_{2}\in\mathbb{F}_p^{*}}\sum_{\gamma_{1}\in\mathbb{F}_p}|T_a(\gamma_{1}^2\gamma_{2}^{-1},\gamma_{1},\gamma_{2})|\\
	=&p^{m-2}-1+\!\!\!\!\!\!\sum_{\substack{(\gamma_{2},\gamma_{1})\in\mathbb{F}_p^{*}\times\mathbb{F}_p\\ \gamma_{2}+b_1\gamma_{1}+b_0\gamma_{1}^2\gamma_{2}^{-1}=0}}\!\!\!\!\!\!|T_a(\gamma_{1}^2\gamma_{2}^{-1},\gamma_{1},\gamma_{2})|\\
	=&p^{m-2}-1+\!\!\!\!\!\!\sum_{\substack{(\gamma_{2},\gamma_{1})\in\mathbb{F}_p^{*}\times\mathbb{F}_p^{*}\\ (\gamma_{2}\gamma_{1}^{-1})^2+b_1(\gamma_{2}\gamma_{1}^{-1})+b_0=0}}\!\!\!\!\!\!|T_a(\gamma_{1}^2\gamma_{2}^{-1},\gamma_{1},\gamma_{2})|\\
	=&p^{m-2}-1.
	\end{align*}$\hfill\Box$
	
	\begin{lemma}\label{l48}		For any integer $m\ge 3$, $a\in\mathbb{F}_{p^{m}}\backslash\mathbb{F}_p$, $\epsilon\in\{-1,1\}$ and $\gamma\in\mathbb{F}_p^{*}$, we have
		
		\begin{align*}
		|{M}_{2,a}(\epsilon,\gamma)|=\begin{cases}
		\frac{p^{m}+\epsilon\cdot\eta_{1}(-\gamma)G_1G_mI_{2}(a)}{2p^3},\quad&\text{~if~} m \text{~is~odd};\\
		 \frac{p^{m}+\epsilon\cdot G_m\big((p-1)(1+\eta_{m}(a)I_1(a))-I_{2}(a)\big)}{2p^3},\quad&\text{~if~} m \text{~is~even} \text{~and~} a\in\mathbb{F}_{p^m}\backslash\mathbb{F}_{p^2};\\
		0, \quad&\text{~if~} m \text{~is~even}\text{~and~}a\in\mathbb{F}_{p^2}\backslash\mathbb{F}_p.
		\end{cases}
		\end{align*}
		
	\end{lemma}

	{\bf Proof.}~By Lemma $\ref{l45}$, one has \begin{align*}
	|{M}_{2,a}(\epsilon,\gamma)|=&L_a(\epsilon,\gamma,0,0)
	=\frac{|T_a(\gamma,0,0)|}{2}+\epsilon\cdot\frac{G_m}{2p^3}\sum\limits_{(z_0,z_1,z_2)\in M(a)}\zeta_p^{-\gamma z_0}\eta_{m}(z_2a^2+z_1a+z_0).
	\end{align*}	
	
	Now by Lemmas \ref{l43}-\ref{l44}, the value of $|{M}_{2,a}(\epsilon,\gamma)|$ is given by the following two cases.
	
	$(1)$ If $m$ is odd, or $m$ is even and $a\in\mathbb{F}_{p^m}\backslash\mathbb{F}_{p^2}$, 
	\begin{align*}
		&|{M}_{2,a}(\epsilon,\gamma)|\\=&\frac{p^{m-3}}{2}+\epsilon\cdot\frac{G_m}{2p^3}\sum\limits_{(z_0,z_1,z_2)\in M(a)}\zeta_p^{-\gamma z_0}\eta_{m}(z_2a^2+z_1a+z_0)\\
	=&\frac{p^{m-3}}{2}+\epsilon\cdot\frac{G_m}{2p^3}\sum\limits_{(z_0,z_1,z_2)\in (\mathbb{F}_p^{3})^{*}}\zeta_p^{-\gamma z_0}\eta_{m}(z_2a^2+z_1a+z_0)\\
	=&\frac{p^{m-3}}{2}+\epsilon\cdot\frac{G_m}{2p^3}\Bigg(\sum\limits_{(z_1,z_2)\in (\mathbb{F}_p^{2})^{*}}\eta_{m}(z_2a^2+z_1a)+
	\sum\limits_{z_0\in\mathbb{F}_p^{*}}\zeta_p^{-\gamma z_0}\eta_{m}(z_0)\sum_{z_1\in\mathbb{F}_p}\sum_{z_2\in\mathbb{F}_p}\eta_{m}(z_2a^2+z_1a+1)\Bigg)\\
	=&\frac{p^{m-3}}{2}+\epsilon\cdot\frac{G_m}{2p^3}\Bigg(\eta_{m}(a)\sum\limits_{(z_1,z_2)\in (\mathbb{F}_p^{2})^{*}}\eta_{m}(z_2a+z_1)+
	I_{2}(a)\sum\limits_{z_0\in\mathbb{F}_p^{*}}\zeta_p^{-\gamma z_0}\eta_{m}(z_0)\Bigg)\\
	=&\frac{p^{m-3}}{2}+\epsilon\cdot\frac{G_m}{2p^3}\bigg(\sum\limits_{z_2\in \mathbb{F}_p^{*}}\eta_{m}(z_2)+\eta_{m}(a)\sum\limits_{z_1\in \mathbb{F}_p^{*}}\eta_{m}(z_1)\sum\limits_{z_2\in \mathbb{F}_p}\eta_{m}(z_2a+1)+
	I_{2}(a)\sum\limits_{z_0\in\mathbb{F}_p^{*}}\zeta_p^{-\gamma z_0}\eta_{m}(z_0)\bigg)\\
	=&\begin{cases}
	\frac{p^{m}+\epsilon\cdot\eta_{1}(-\gamma)G_1G_mI_{2}(a)}{2p^3},\quad & \text{if~} m \text{~is~odd};\\
	\frac{p^{m}+\epsilon\cdot G_m\big((p-1)(1+\eta_{m}(a)I_1(a))-I_{2}(a)\big)}{2p^3},\quad& \text{if~} m \text{~is~even}\text{~and~}a\in\mathbb{F}_{p^m}\backslash\mathbb{F}_{p^2}.
	\end{cases}
	\end{align*}
	
	$(2)$ If $m$ is even and $a\in\mathbb{F}_{p^2}\backslash\mathbb{F}_p$, then there exist some $b_1\in\mathbb{F}_p$ and $b_0\in\mathbb{F}_p^{*}$ such that $a^2+b_1a+b_0=\!0$, thus
	\begin{align*}
		&|{M}_{2,a}(\epsilon,\gamma)|\\=&\epsilon\cdot\frac{G_m}{2p^3}\sum\limits_{(z_0,z_1,z_2)\in M(a)}\zeta_p^{-\gamma z_0}\eta_{m}(z_2a^2+z_1a+z_0)\\
	=&\epsilon\cdot\frac{G_m}{2p^3}\sum\limits_{(z_0,z_1,z_2)\in (\mathbb{F}_p^{3})^*}\zeta_p^{-\gamma z_0}\eta_{m}(z_2a^2+z_1a+z_0)-\sum\limits_{\substack{(z_0,z_1,z_2)\in \mathbb{F}_p^{3}\\(z_1,z_0)=(z_2b_1,z_2b_0)}}\zeta_p^{-\gamma z_0}\eta_{m}(z_2a^2+z_1a+z_0)\\
	=&\epsilon\cdot\frac{G_m}{2p^3}\sum\limits_{(z_0,z_1,z_2)\in (\mathbb{F}_p^{3})^*}\zeta_p^{-\gamma z_0}\eta_{m}(z_2a^2+z_1a+z_0)-\sum\limits_{z_2\in \mathbb{F}_p}\zeta_p^{\gamma b_0 z_2}\eta_{m}(z_2a^2+z_2b_1a+z_2b_0)\\
	=&\epsilon\cdot\frac{G_m}{2p^3}\sum\limits_{(z_0,z_1,z_2)\in (\mathbb{F}_p^{3})^*}\zeta_p^{-\gamma z_0}\eta_{m}(z_2a^2+z_1a+z_0)\\
	=&\epsilon\cdot\frac{G_m}{2p^3}\Big(\sum\limits_{(z_1,z_2)\in (\mathbb{F}_p^{2})^*}\eta_{m}(z_2a^2+z_1a)-\sum\limits_{(z_1,z_2)\in (\mathbb{F}_p^{2})}\eta_{m}(z_2a^2+z_1a+1)\Big)\\
	=&\epsilon\cdot\frac{G_m}{2p^3}\bigg(\eta_{m}(a)\Big(\sum\limits_{z_2\in \mathbb{F}_p^{*}}\eta_{m}(z_2a)+\sum\limits_{z_1\in \mathbb{F}_p^{*}}\sum\limits_{z_2\in \mathbb{F}_p}\eta_{m}(z_2a+z_1)\Big)-\sum_{z_2\in\mathbb{F}_{p}}\sum_{z_1\in\mathbb{F}_{p}}\eta_{m}(z_2a^2+z_1a+1)\bigg)\\
	=&\frac{\epsilon\cdot G_m\Big((p-1)\big(\eta_{m}(a)I_1(a)+1\big)-I_{2}(a)\Big)}{2p^3},
	\end{align*}
	and then by	$|{M}_{2,a}(\epsilon,\gamma)|\ge 0$, we know that   
	\begin{align*}
		\frac{\epsilon\cdot G_m\Big((p-1)\big(\eta_{m}(a)I_1(a)+1\big)-I_{2}(a)\Big)}{2p^3}\ge 0\quad \big(\epsilon\in\{-1,1\}\big), 
	\end{align*}
	it implies that
	\begin{align}\label{l481}
		(p-1)\Big(\eta_{m}(a)I_1(a)+1\Big)-I_{2}(a)=0.
	\end{align}
	
	So far, we complete the proof. $\hfill\Box$
	
	\begin{remark}\label{r1}
		By $(\ref{l520})$ and $(\ref{l481})$, one has
		\begin{align}\label{l524}
		(p-1)\Big(\eta_{m}(a)I_1(a)+1\Big)-I_{2}(a)=(p-1)(\eta_{m}(a)-1)(I_1(a)-1)=0.
		\end{align}	
Particularly, if $\eta_m(a)=-1$, then we have 
		\begin{align}\label{l521}
		I_1(a)=1\quad\text{~and~}\quad I_{2}(a)=0.
		\end{align}
	\end{remark}
	\begin{lemma}\label{l49}
			For any integer $m\ge 3$ and $a\in\mathbb{F}_{p^{m}}\backslash\mathbb{F}_p$, we have
		\begin{align*}
	|{M}_{3,a}(\epsilon)|=\begin{cases}
	\frac{(p-1)\Big(p^m+\epsilon\cdot G_m\big((p-1)-\eta_{m}(a)I_1(a)\big)\Big)}{2p^2},\quad&\text{~if~}m\text{~is even};\\
		\frac{(p-1)p^{m-2}}{2}, \quad&\text{~if~}m\text{~is odd~}.
		\end{cases}
		\end{align*}
	\end{lemma}

	{\bf Proof.}~By calculating directly, we have
	
	\begin{align*}
	M_{3,a}(\epsilon)=\bigcup_{\gamma_{1}\in\mathbb{F}_p^{*}}\bigcup_{\gamma_{0}\in\mathbb{F}_p}L_a(\epsilon,\gamma_{0},\gamma_{1},0).
	\end{align*} 
	And then\begin{align*}
		|{M}_{3,a}(\epsilon)|
		=&\sum\limits_{\gamma_{1}\in\mathbb{F}_p^{*}}\sum\limits_{\gamma_{0}\in\mathbb{F}_p}\frac{|T_a(\gamma_{0},\gamma_{1},0)|}{2}+\epsilon\cdot\frac{G_m}{2p^3}
	\sum_{\gamma_{1}\in\mathbb{F}_p^{*}}\sum_{\gamma_{0}\in\mathbb{F}_p}\sum\limits_{(z_0,z_1,z_2)\in M(a)}\zeta_p^{-\gamma_{1}z_1-\gamma_{0}z_0}\eta_{m}(z_2a^2+z_1a+z_0)\\
	=&\sum\limits_{\gamma_{1}\in\mathbb{F}_p^{*}}\sum\limits_{\gamma_{0}\in\mathbb{F}_p}\frac{|T_a(\gamma_{0},\gamma_{1},0)|}{2}+\epsilon\cdot\frac{G_m}{2p^2}\eta_m(a)\sum_{\gamma_{1}\in\mathbb{F}_p^{*}}\sum\limits_{(0,z_1,z_2)\in M(a)}\zeta_p^{-\gamma_{1}z_1}\eta_{m}(z_2a+z_1).
	\end{align*}
	Now by Lemma \ref{l43}, we know that $(0,z_1,z_2)\in M(a)$ if and only if $(z_1,z_2)\in\big(\mathbb{F}_q^2\big)^*$, and then by Lemma \ref{l44}, $|{M}_{3,a}(\epsilon)|$ can be obtained in the following two cases.
	
	$(1)$~If $m$ is odd, or $m$ is even and $a\in\mathbb{F}_{p^m}\backslash\mathbb{F}_{p^2}$, 
	
{\small	\begin{align*}
|{M}_{3,a}(\epsilon)|=&
	\frac{(p-1)p^{m-2}}{2}+\epsilon\!\cdot\!\frac{G_m}{2p^2}\eta_m(a)\sum_{\gamma_{1}\in\mathbb{F}_p^{*}}\sum\limits_{(z_1,z_2)\in(\mathbb{F}_p^{2})^{*}}\zeta_p^{-\gamma_{1}z_1}\eta_{m}(z_2a+z_1)\\
	=&
	\frac{(p-1)p^{m-2}}{2}+\epsilon\!\cdot\!\frac{G_m}{2p^2}\eta_m(a)\Big(\eta_{m}(a)(p-1)\sum_{z_2\in\mathbb{F}_p^*}\eta_{m}(z_2)-\sum_{z_1\in\mathbb{F}_p^{*}}\eta_{m}(z_1)\sum\limits_{z_2\in\mathbb{F}_p}\eta_{m}(z_2a+1)\Big)\\
	=&\begin{cases}	
	\frac{(p-1)p^{m-2}}{2}, \quad&\text{~if~} m \text{~is~odd};\\
	\frac{(p-1)\Big(p^m+\epsilon\cdot G_m\big((p-1)-\eta_{m}(a)I_1(a)\big)\Big)}{2p^2},\quad&\text{~if~} m \text{~is~even}\text{~and~} a\in\mathbb{F}_{p^m}\backslash\mathbb{F}_{p^2}.
	\end{cases}
	\end{align*}}
	
	$(2)$~If $m$ is even and $a\in\mathbb{F}_{p^2}\backslash\mathbb{F}_p$, then there exist some $b_1\in\mathbb{F}_p$ and $b_0\in\mathbb{F}_p^{*}$ such that $a^2+b_1a+b_0=\!0$, thus
{\small	\begin{align*}
	|{M}_{3,a}(\epsilon)|=&\sum\limits_{\gamma_{1}\in\mathbb{F}_p^{*}}\frac{|T_a(-b_0^{-1}b_1\gamma_1,\gamma_{1},0)|}{2}+\epsilon\cdot\frac{G_m}{2p^2}\eta_m(a)\sum_{\gamma_{1}\in\mathbb{F}_p^{*}}\sum\limits_{(0,z_1,z_2)\in M(a)}\zeta_p^{-\gamma_{1}z_1}\eta_{m}(z_2a+z_1)\\
	=&\frac{(p-1)p^{m-2}}{2}+\epsilon\cdot\frac{G_m}{2p^2}\eta_m(a)\sum_{\gamma_{1}\in\mathbb{F}_p^{*}}\sum\limits_{(z_1,z_2)\in (\mathbb{F}_p^{2})^{*}}\zeta_p^{-\gamma_{1}z_1}\eta_{m}(z_2a+z_1)\\
	=&\frac{(p-1)p^{m-2}}{2}+\epsilon\!\cdot\!\frac{G_m}{2p^2}\eta_m(a)\Big(\sum_{\gamma_{1}\in\mathbb{F}_p^{*}}\sum\limits_{z_2\in \mathbb{F}_p^{*}}\!\eta_{m}(z_2a)+\!\sum_{\gamma_{1}\in\mathbb{F}_p^{*}}\sum_{z_1\in\mathbb{F}_p^{*}}\zeta_p^{-\gamma_{1}z_1}\!\sum\limits_{z_2\in \mathbb{F}_p}\eta_{m}(z_2a+z_1)\Big)\\
	=&\frac{(p-1)\Big(p^m+\epsilon\cdot G_m\big((p-1)-\eta_m(a)I_1(a)\big)\Big)}{2p^2}.
	\end{align*}}
$\hfill\Box$
	
	\begin{lemma}\label{l410}
			For any integer $m\ge 3$, $a\in\mathbb{F}_{p^m}\backslash\mathbb{F}_p$, $\epsilon\in\{-1,1\}$ and $\gamma\in\mathbb{F}_p^{*}$,  we have
		
		\begin{align*}
		|\mathrm{M}_{4,a}(\epsilon,\gamma)|=\begin{cases}
		\frac{(p-1)p^{m+1}+\epsilon\cdot\eta_{1}(-1)G_1G_m\big( I_{2}(a)-p\big)}{2p^3},&\text{~if~} m \text{~is~odd};\\
		\frac{(p-1)p^{m}+\epsilon\cdot\eta_{1}(-\gamma) G_m\big(p^2-I_{2}(a)\big)}{2p^2},&\text{~if~} m \text{~is~even}\text{~and~}a\in\mathbb{F}_{p^m}\backslash\mathbb{F}_{p^2};\\	
		\frac{p^{m+1}+\epsilon\cdot\eta_{1}(-\gamma) G_m\big(p^2-I_{2}(a)\big)}{2p^2}, &\text{~if~} m \text{~is~even}\text{~and~}a\in\mathbb{F}_{p^2}\backslash\mathbb{F}_{p}.
		\end{cases}
		\end{align*}
	\end{lemma}

	{\bf Proof.} By calculating directly, we have 
{\small	\begin{align*} 
		&\mathrm{M}_{4,a}(\epsilon,\gamma)\\=&\bigcup\limits_{\epsilon_1\in\{-1,1\}}\bigcup\limits_{\delta\in\{-1,1\}}\bigcup\limits_{\epsilon_1\delta=\epsilon}\!\left\{\!b\in\mathbb{F}_{p^m}^{*}\big|\eta_{m}(b)=\epsilon_1,\eta_{1}\big(\mathrm{Tr}(a^2b^{-1})\big)=\delta~\!\text{and}~\!\frac{\mathrm{Tr}(a^2b^{-1})\mathrm{Tr}(b^{-1})-\mathrm{Tr}(ab^{-1})^2}{\mathrm{Tr}(a^2b^{-1})}=\gamma\!\right\}\\
		=&\bigcup\limits_{\epsilon_1\in\{-1,1\}}\bigcup\limits_{\gamma_{1}\in\mathbb{F}_p}\bigcup\limits_{\substack{\gamma_{2}\in\mathbb{F}_p\\\epsilon_1\eta_{1}(\gamma_{2})=\epsilon}}\!\bigg\{\!b\in\mathbb{F}_{p^m}^{*}|\eta_{m}(b)=\epsilon_1,\mathrm{Tr}\big(ab^{-1}\big)=\gamma_{1},\mathrm{Tr}({a^2}b^{-1})=\gamma_{2}~\!\text{and}~\!\mathrm{Tr}\big(b^{-1}\big)=\gamma_{1}^2\gamma_{2}^{-1}+\gamma\!\bigg\}\\=&\bigcup\limits_{\epsilon_1\in\{-1,1\}}\bigcup\limits_{\gamma_{1}\in\mathbb{F}_p}\bigcup\limits_{\substack{\gamma_{2}\in\mathbb{F}_p\\\eta_{1}(\gamma_{2})=\epsilon\epsilon_1}}L_a\Big(\epsilon_1,\gamma_{1}^2\gamma_{2}^{-1}+\gamma,\gamma_{1},\gamma_{2}\Big),
	\end{align*}}
	thus {\small\begin{align}\label{l4101}\begin{aligned}
			&|\mathrm{M}_{4,a}(\epsilon,\gamma)|\\
		=&\Bigg|\bigcup\limits_{\epsilon_1\in\{-1,1\}}\bigcup\limits_{\gamma_{1}\in\mathbb{F}_p}\bigcup\limits_{\substack{\gamma_{2}\in\mathbb{F}_p\\\eta_{1}(\gamma_{2})=\epsilon\epsilon_1}}L_a\Big(\epsilon_1,\gamma_{1}^2\gamma_{2}^{-1}+\gamma,\gamma_{1},\gamma_{2}\Big)\Bigg|\\
		=&\sum\limits_{\epsilon_1\in\{-1,1\}}\sum_{\gamma_{1}\in\mathbb{F}_p}\sum\limits_{\substack{\gamma_{2}\in\mathbb{F}_p\\\eta_{1}(\gamma_{2})=\epsilon\epsilon_1}}\frac{|T_a(\gamma_{1}^2\gamma_{2}^{-1}+\gamma,\gamma_{1},\gamma_{2})|}{2}\\
		&+\frac{ G_m}{2p^3}\sum_{\gamma_{1}\in\mathbb{F}_p}\sum\limits_{\substack{\gamma_{2}\in\mathbb{F}_p\\\eta_{1}(\gamma_{2})=\epsilon}}\sum\limits_{(z_0,z_1,z_2)\in M(a)}\zeta_p^{-\gamma_{2}z_2-\gamma_{1}z_1-\Big(\gamma_{1}^2\gamma_{2}^{-1}+\gamma\Big) z_0}\eta_{m}(z_2a^2+z_1a+z_0)\\
		&-\frac{ G_m}{2p^3}\sum_{\gamma_{1}\in\mathbb{F}_p}\sum\limits_{\substack{\gamma_{2}\in\mathbb{F}_p\\\eta_{1}(\gamma_{2})=-\epsilon}}\sum\limits_{(z_0,z_1,z_2)\in M(a)}\zeta_p^{-\gamma_{2}z_2-\gamma_{1}z_1-\Big(\gamma_{1}^2\gamma_{2}^{-1}+\gamma\Big) z_0}\eta_{m}(z_2a^2+z_1a+z_0)\\
		=&\frac{1}{2}\sum_{\epsilon_1\in\{1,-1\}}\sum_{\gamma_{1}\in\mathbb{F}_p}\sum\limits_{\substack{\gamma_{2}\in\mathbb{F}_p\\\eta_{1}(\gamma_{2})=\epsilon_1}}|T_a(\gamma_{1}^2\gamma_{2}^{-1}+\gamma,\gamma_{1},\gamma_{2})|\\
		&+\frac{ G_m}{2p^3}\sum_{\gamma_{1}\in\mathbb{F}_p}\sum\limits_{\gamma_{2}\in\mathbb{F}_{p}^{*}}\frac{1+\epsilon\eta_{1}(\gamma_{2})}{2}\!\!\!\!\!\sum\limits_{(z_0,z_1,z_2)\in M(a)}\!\!\!\!\!\zeta_p^{-\gamma_{2}z_2-\gamma_{1}z_1-\Big(\gamma_{1}^2\gamma_{2}^{-1}+\gamma\Big) z_0}\eta_{m}(z_2a^2+z_1a+z_0)\\
		&-\frac{ G_m}{2p^3}\sum_{\gamma_{1}\in\mathbb{F}_p}\sum\limits_{\gamma_{2}\in\mathbb{F}_{p}^{*}}\frac{1-\epsilon\eta_{1}(\gamma_{2})}{2}\!\!\!\!\!\sum\limits_{(z_0,z_1,z_2)\in M(a)}\!\!\!\!\!\zeta_p^{-\gamma_{2}z_2-\gamma_{1}z_1-\Big(\gamma_{1}^2\gamma_{2}^{-1}+\gamma\Big) z_0}\eta_{m}(z_2a^2+z_1a+z_0)\\
		=&\begin{cases}
		\frac{(p-1)p^{m-2}}{2}+\epsilon\cdot\frac{G_m}{2p^3}E_{1,a}(\gamma),\quad&\text{~if~}a\in\mathbb{F}_{p^m}\backslash\mathbb{F}_{p^2};\\
		\frac{p^{m-1}}{2}+\epsilon\cdot\frac{G_m}{2p^3}\big(E_{1,a}(\gamma)-E_{2,a}(\gamma)\big),\quad&\text{~if~}a\in\mathbb{F}_{p^2}\backslash\mathbb{F}_p.
		\end{cases}
		\end{aligned}
	\end{align}}

	Now by $(\ref{l4101})$ and Lemmas \ref{l46}-\ref{l47}, we complete the proof.
	
	$\hfill\Box$
	\subsection{Proofs for Theorems \ref{t1}-\ref{t3}}
	{\bf The proof for Theorem \ref{t1}}. 
	
	For any odd $m$, by Lemma \ref{l41} we know that $\mathcal{C}_{D_a}$ is with length $n=p^{m-2}$. And by Lemma \ref{l42} we can calculate the complete weight enumerator for $\mathcal{C}_{D_a}$ in the following $4$ cases.\\
	
	{\bf Case 1.}  If $\mathrm{Tr}(a^2b^{-1})=\mathrm{Tr}(ab^{-1})=\mathrm{Tr}(b^{-1})=0, \text{~or~}\mathrm{Tr}(a^2b^{-1})\neq 0 \text{~and~} \mathrm{Tr}(a^2b^{-1})\mathrm{Tr}(b^{-1})-\mathrm{Tr}(ab^{-1})^2=0$, then
	\begin{align*}
	|N_{a}(b,\rho)|=p^{m-3}.
	\end{align*} 
	Now by Lemma \ref{l47}, the frequency is $$|M_{1,a}|=p^{m-1}-p^{m-2}+p^{m-3}-1.$$

	{\bf Case 2.} For $\epsilon\in\{1,-1\}$ and $\gamma\in\mathbb{F}_p^{*}$, if $\eta_m(b)=\epsilon$, $\mathrm{Tr}(a^2b^{-1})=\mathrm{Tr}(ab^{-1})=0$ and $\mathrm{Tr}(b^{-1})=\gamma^{-1}$, then 
	\begin{align*}
	|N_{a}(b,\rho)|=\begin{cases}
	p^{m-3}+\epsilon\cdot\frac{\eta_{1}(-\gamma)(p-1) G_mG_1}{p^2},\quad&\text{~if~} \rho=\gamma;\\
	p^{m-3}-\epsilon\cdot\frac{\eta_{1}(-\gamma)G_mG_1}{p^2},\quad&\text{~if~} \rho\neq\gamma.
	\end{cases}
	\end{align*} 
	Now by Lemma \ref{l48}, the frequency is $$|M_{2,a}(\epsilon,\gamma)|=\frac{p^{m}+\epsilon\cdot\eta_{1}(-\gamma)G_1G_mI_{2}(a)}{2p^3}.$$

	{\bf Case 3.} For $\epsilon\in\{1,-1\}$, if $\eta_m(b)=\epsilon$, $\mathrm{Tr}(a^2b^{-1})=0$ and $\mathrm{Tr}(ab^{-1})\neq 0$, then
	\begin{align*}
	|N_{a}(b,\rho)|=\begin{cases}
	p^{m-3},\quad &\text{~if~}\rho=0;\\
	p^{m-3}-\epsilon\cdot \frac{\eta_{1}(-\rho)G_mG_1}{p^2},\quad &\text{~if~}\rho\neq 0.
	\end{cases}
	\end{align*} 
	Now by Lemma \ref{l49}, the frequency is $$|M_{3,a}(\epsilon)|=\frac{(p-1)p^{m-2}}{2}.$$

	{\bf Case 4.} For $\epsilon\in\{1,-1\}$ and $\gamma\in\mathbb{F}_p^{*}$, if $\eta_m(b)\eta_{1}\big(\mathrm{Tr}(a^2b^{-1})\big)=\epsilon$, $\mathrm{Tr}(a^2b^{-1})\neq 0$ and $\frac{\mathrm{Tr}(a^2b^{-1})\mathrm{Tr}(b^{-1})-\mathrm{Tr}(ab^{-1})^2}{\mathrm{Tr}(a^2b^{-1})}=\gamma^{-1}$, then
	\begin{align*}
	|N_{a}(b,\rho)|=\begin{cases}
	p^{m-3},\quad& \text{~if~}\rho=\gamma;\\
	p^{m-3}+\epsilon\cdot\frac{\eta_{1}\big(\rho\gamma^{-1}-1\big)G_mG_1}{p^2},\quad&\text{~if~} \rho\neq\gamma.
	\end{cases}
	\end{align*}	
	Now by Lemma \ref{l410}, the frequency is $$|M_{4,a}(\epsilon,\gamma^{-1})|=\frac{(p-1)p^{m+1}+\epsilon\cdot\eta_{1}(-1) G_1G_m\big(I_{2}(a)-p\big)}{2p^3}.$$
	
	By the above disscussions and Lemma \ref{l22}, we can obtain the complete weight enumerator for $\mathcal{C}_{D_a}$ directly. 
	Now, in order to give the weight distribution for $\mathcal{C}_{D_a}$, we need the value of $|N_{a}(b,0)|$ and its frequency accroding to the following three cases.\\
	
	$(1)$ If $\mathrm{Tr}(a^2b^{-1})=\mathrm{Tr}(ab^{-1})=\mathrm{Tr}(b^{-1})=0, \text{~or~}\mathrm{Tr}(a^2b^{-1})\neq 0 \text{~and~} \mathrm{Tr}(a^2b^{-1})\mathrm{Tr}(b^{-1})-\mathrm{Tr}(ab^{-1})^2=0$, or $\mathrm{Tr}(a^2b^{-1})=0$ and $\mathrm{Tr}(ab^{-1})\neq 0$, then
		\begin{align*}
	|N_{a}(b,0)|=p^{m-3},
	\end{align*} and the frequency is $$|M_{1,a}|+\sum_{\epsilon\in\{-1,1\}}|M_{3,a}(\epsilon)|=2(p-1)p^{m-2}+p^{m-3}-1.$$
	
	$(2)$ For $\epsilon\in\{1,-1\}$ and $\gamma\in\mathbb{F}_p^{*}$, if $\eta_m(b)=\epsilon$, $\mathrm{Tr}(a^2b^{-1})=\mathrm{Tr}(ab^{-1})=0$ and $\mathrm{Tr}(b^{-1})=\gamma^{-1}$ with $\epsilon\cdot\eta_{1}(\gamma)=1$, or $\eta_m(b)\eta_{1}\big(\mathrm{Tr}(a^2b^{-1})\big)=-1$  and $\frac{\mathrm{Tr}(a^2b^{-1})\mathrm{Tr}(b^{-1})-\mathrm{Tr}(ab^{-1})^2}{\mathrm{Tr}(a^2b^{-1})}=\gamma^{-1}$, then
		\begin{align*}
	|N_{a}(b,0)|=
	p^{m-3}-\frac{\eta_{1}(-1)G_mG_1}{p^2},
	\end{align*}
	and the frequency is{\small \begin{align*}
	&\sum_{\substack{\gamma\in\mathbb{F}_p\\\eta_{1}(\gamma)=1}}|M_{2,a}(1,\gamma)|+\sum_{\substack{\gamma\in\mathbb{F}_p\\\eta_{1}(\gamma)=-1}}|M_{2,a}(-1,\gamma)|+\sum_{\gamma\in\mathbb{F}_p}|M_{4,a}(-1,\gamma^{-1})|\\
	=&\frac{(p-1)\big(p^{m}+\eta_{1}(-1)G_1G_mI_{2}(a)\big)}{2p^3}+\frac{(p-1)\Big((p-1)p^{m+1}-\eta_{1}(-1)G_1G_m\big(I_{2}(a)-p\big)\Big)}{2p^3}\\
	=&\frac{(p-1)\big((p-1)p^{m}+p^{m-1}+\eta_{1}(-1)G_1G_m\big)}{2p^2}.
	\end{align*}}
	
		$(3)$ For $\epsilon\in\{1,-1\}$ and $\gamma\in\mathbb{F}_p^{*}$, if $\eta_m(b)=\epsilon$, $\mathrm{Tr}(a^2b^{-1})=\mathrm{Tr}(ab^{-1})=0$ and $\mathrm{Tr}(b^{-1})=\gamma^{-1}$ with $\epsilon\cdot\eta_{1}(\gamma)=-1$, or $\eta_m(b)\eta_{1}\big(\mathrm{Tr}(a^2b^{-1})\big)=1$,  and $\frac{\mathrm{Tr}(a^2b^{-1})\mathrm{Tr}(b^{-1})-\mathrm{Tr}(ab^{-1})^2}{\mathrm{Tr}(a^2b^{-1})}=\gamma^{-1}$, then
	\begin{align*}
	|N_{a}(b,0)|=
	p^{m-3}+\frac{\eta_{1}(-1)G_mG_1}{p^2},
	\end{align*}
	and the frequency is{\small \begin{align*}
	&\sum_{\substack{\gamma\in\mathbb{F}_p\\\eta_{1}(\gamma)=-1}}|M_{2,a}(1,\gamma)|+\sum_{\substack{\gamma\in\mathbb{F}_p\\\eta_{1}(\gamma)=1}}|M_{2,a}(-1,\gamma)|+\sum_{\gamma\in\mathbb{F}_p}|M_{4,a}(1,\gamma^{-1})|\\
	=&\frac{(p-1)\big(p^{m}-\eta_{1}(-1)G_1G_mI_{2}(a)\big)}{2p^3}+\frac{(p-1)\Big((p-1)p^{m+1}+\eta_{1}(-1)G_1G_m\big(I_{2}(a)-p\big)\Big)}{2p^3}\\
	=&\frac{(p-1)\big((p-1)p^{m}+p^{m-1}-\eta_{1}(-1)G_1G_m\big)}{2p^2}.
	\end{align*}	}

	Note that the weight for any codeword  $\mathbf{c}(b)\in\mathcal{C}_{D_a}$ is $p^{m-2}-N_{a}(b,0)$, hence by the above discussions  and Lemma \ref{l22}, we can get the weight distribution for $\mathcal{C}_{D_a}$. $\hfill\Box$\\
	
		{\bf Proofs for Theorems \ref{t2}-\ref{t3}}. 
		
		For any even $m$, by Lemma \ref{l41} we know that $\mathcal{C}_{D_a}$ is with length $n=p^{m-2}$. And by Lemma \ref{l42} we can calculate the complete weight enumerator for $\mathcal{C}_{D_a}$ depending on the following $4$ cases.
	
	{\bf Case 1.}  If $\mathrm{Tr}(a^2b^{-1})=\mathrm{Tr}(ab^{-1})=\mathrm{Tr}(b^{-1})=0, \text{~or~}\mathrm{Tr}(a^2b^{-1})\neq 0 \text{~and~} \mathrm{Tr}(a^2b^{-1})\mathrm{Tr}(b^{-1})-\mathrm{Tr}(ab^{-1})^2=0$, then
	\begin{align*}
	|N_{a}(b,\rho)|=p^{m-3}.
	\end{align*} 
	Now by Lemma \ref{l47}, the frequency is $$|M_{1,a}|=\begin{cases}
	p^{m-1}-p^{m-2}+p^{m-3}-1,\quad&\text{~if~}a\in\mathbb{F}_{p^{m}}\backslash\mathbb{F}_{p^2};	\\
	p^{m-2}-1,\quad &\text{~if~}a\in\mathbb{F}_{p^2}\backslash\mathbb{F}_p.
	\end{cases}$$
	
	{\bf Case 2.} For $\epsilon\in\{1,-1\}$ and $\gamma\in\mathbb{F}_p^{*}$, if $\eta_m(b)=\epsilon$, $\mathrm{Tr}(a^2b^{-1})=\mathrm{Tr}(ab^{-1})=0$ and $\mathrm{Tr}(b^{-1})=\gamma^{-1}$, then 
	\begin{align*}
	|N_{a}(b,\rho)|=\begin{cases}
	p^{m-3},\quad&\text{~if~} \rho=\gamma;\\
	p^{m-3}+\epsilon\cdot\frac{\eta_{1}(1-\rho\gamma^{-1})G_m}{p},\quad&\text{~if~} \rho\neq\gamma.
	\end{cases}
	\end{align*} 
	Now by Lemma \ref{l48}, the frequency is 
	\begin{align*}
	|{M}_{2,a}(\epsilon,\gamma)|=\begin{cases}
	\frac{p^{m}+\epsilon\cdot G_m\big((p-1)(1+\eta_{m}(a)I_1(a))-I_{2}(a)\big)}{2p^3},\quad&\text{~if~} a\in\mathbb{F}_{p^m}\backslash\mathbb{F}_{p^2};\\
0,\quad&\text{~if~}a\in\mathbb{F}_{p^2}\backslash\mathbb{F}_p.
	\end{cases}
	\end{align*}

	{\bf Case 3.} For any given $\epsilon\in\{1,-1\}$, if $\eta_m(b)=\epsilon$, $\mathrm{Tr}(a^2b^{-1})=0$ and $\mathrm{Tr}(ab^{-1})\neq 0$, then
	\begin{align*}
	|N_{a}(b,\rho)|=\begin{cases}
	p^{m-3}+\epsilon\cdot\frac{(p-1)G_m}{p^2},\quad &\text{~if~}\rho=0;\\
	p^{m-3}-\epsilon\cdot \frac{G_m}{p^2},\quad &\text{~if~}\rho\neq 0.
	\end{cases}
	\end{align*} 
	Now by Lemma \ref{l49}, the frequency is 	\begin{align*}
	|{M}_{3,a}(\epsilon)|=		\frac{(p-1)\big(p^m+\epsilon\cdot G_m\big((p-1)-\eta_{m}(a)I_1(a)\big)\big)}{2p^2}.
	\end{align*}
	
	{\bf Case 4.} For $\epsilon\in\{1,-1\}$ and $\gamma\in\mathbb{F}_p^{*}$, if $\eta_m(b)\eta_{1}\big(\mathrm{Tr}(a^2b^{-1})\big)=\epsilon$ and $\frac{\mathrm{Tr}(a^2b^{-1})\mathrm{Tr}(b^{-1})-\mathrm{Tr}(ab^{-1})^2}{\mathrm{Tr}(a^2b^{-1})}=\gamma^{-1}$, then
	\begin{align*}
	|N_{a}(b,\rho)|=\begin{cases}
	p^{m-3}+\epsilon\cdot\frac{\eta_{1}\big(-\gamma\big)(p-1)G_m}{p^2},\quad&\text{~if~} \rho=\gamma;\\
	p^{m-3}-\epsilon\cdot\frac{\eta_{1}\big(-\gamma\big)G_m}{p^2},\quad&\text{~if~} \rho\neq\gamma.
	\end{cases}
	\end{align*}\\	
	Now by Lemma \ref{l410}, the frequency is\begin{align*}
	|\mathrm{M}_{4,a}(\epsilon,\gamma)|=\begin{cases}
	\frac{(p-1)p^{m}+\epsilon\cdot\eta_{1}(-\gamma) G_m\big(p^2-I_{2}(a)\big)}{2p^2},&\text{~if~}a\in\mathbb{F}_{p^m}\backslash\mathbb{F}_{p^2};\\	
	\frac{p^{m+1}+\epsilon\cdot\eta_{1}(-\gamma) G_m\big(p^2-I_{2}(a)\big)}{2p^2}, &\text{~if~}a\in\mathbb{F}_{p^2}\backslash\mathbb{F}_{p}.
	\end{cases}
	\end{align*}
	
	By the above disscussions and Lemma \ref{l22}, the complete weight enumerator for $\mathcal{C}_{D_a}$ is obtained directly. 
	
	And now, in order to give the weight distribution for $\mathcal{C}_{D_a}$, we need the value of $|N_{a}(b,0)|$ and its frequency accroding to the following seven cases.
	
	$(1)$ If $\mathrm{Tr}(a^2b^{-1})=\mathrm{Tr}(ab^{-1})=\mathrm{Tr}(b^{-1})=0, \text{~or~}\mathrm{Tr}(a^2b^{-1})\neq 0 \text{~and~} \mathrm{Tr}(a^2b^{-1})\mathrm{Tr}(b^{-1})-\mathrm{Tr}(ab^{-1})^2=0$,  then
	\begin{align*}
	|N_{a}(b,0)|=p^{m-3},
	\end{align*} and the frequency is $$|M_{1,a}|=\begin{cases}
	p^{m-1}-p^{m-2}+p^{m-3}-1,\quad&\text{~if~}a\in\mathbb{F}_{p^{m}}\backslash\mathbb{F}_{p^2};	\\
	p^{m-2}-1,\quad &\text{~if~}a\in\mathbb{F}_{p^2}\backslash\mathbb{F}_p.
	\end{cases}$$
	
	$(2)$   If $\eta_m(b)=1$, $\mathrm{Tr}(a^2b^{-1})=\mathrm{Tr}(ab^{-1})=0$ and $\mathrm{Tr}(b^{-1})\neq 0$, then 
	\begin{align*}
	|N_{a}(b,0)|=
	p^{m-3}+ \frac{G_m}{p},
	\end{align*} and the frequency is 
	\begin{align*}
	\sum_{\gamma\in\mathbb{F}_p^{*}}|{M}_{2,a}(1,\gamma)|=\begin{cases}
	\frac{(p-1)\Big(p^{m}+ G_m\big((p-1)(1+\eta_{m}(a)I_1(a))-I_{2}(a)\big)\Big)}{2p^3},\quad&\text{~if~} a\in\mathbb{F}_{p^m}\backslash\mathbb{F}_{p^2};\\
	0,\quad&\text{~if~}a\in\mathbb{F}_{p^2}\backslash\mathbb{F}_p.
	\end{cases}
	\end{align*}
	
	$(3)$   If $\eta_m(b)=-1$, $\mathrm{Tr}(a^2b^{-1})=\mathrm{Tr}(ab^{-1})=0$ and $\mathrm{Tr}(b^{-1})\neq 0$, then 
	\begin{align*}
	|N_{a}(b,0)|=
	p^{m-3}- \frac{G_m}{p},
	\end{align*} and the frequency is 
		\begin{align*}
	\sum_{\gamma\in\mathbb{F}_p^{*}}|{M}_{2,a}(-1,\gamma)|=\begin{cases}
	\frac{(p-1)\Big(p^{m}- G_m\big((p-1)(1+\eta_{m}(a)I_1(a))-I_{2}(a)\big)\Big)}{2p^3},\quad&\text{~if~} a\in\mathbb{F}_{p^m}\backslash\mathbb{F}_{p^2};\\
	0,\quad&\text{~if~}a\in\mathbb{F}_{p^2}\backslash\mathbb{F}_p.
	\end{cases}
	\end{align*}

	$(4)$  If $\eta_m(b)=1$, $\mathrm{Tr}(a^2b^{-1})=0$ and $\mathrm{Tr}(ab^{-1})\neq 0$, then
	\begin{align*}
	|N_{a}(b,0)|=
	p^{m-3}+\frac{(p-1)G_m}{p^2},
	\end{align*} and  the frequency is 		\begin{align*}
	|{M}_{3,a}(1)|=		\frac{(p-1)\big(p^m+G_m\big((p-1)-\eta_{m}(a)I_1(a)\big)\big)}{2p^2}.
	\end{align*}
	
	$(5)$  If $\eta_m(b)=-1$, $\mathrm{Tr}(a^2b^{-1})=0$ and $\mathrm{Tr}(ab^{-1})\neq 0$, then
	\begin{align*}
	|N_{a}(b,0)|=
	p^{m-3}-\frac{(p-1)G_m}{p^2},
	\end{align*}  and the frequency is 		\begin{align*}
	|{M}_{3,a}(-1)|=		\frac{(p-1)\big(p^m- G_m\big((p-1)-\eta_{m}(a)I_1(a)\big)\big)}{2p^2}.
	\end{align*}
	
	$(6)$  If  $\eta_m(b)\eta_{1}\big(\mathrm{Tr}(a^2b^{-1})\big)\eta_{1}\Big(-\frac{\mathrm{Tr}(a^2b^{-1})\mathrm{Tr}(b^{-1})-\mathrm{Tr}(ab^{-1})^2}{\mathrm{Tr}(a^2b^{-1})}\Big)=1$, then
		\begin{align*}
	|N_{a}(b,0)|=
	p^{m-3}+\frac{G_m}{p^2},
	\end{align*} and the frequency is \begin{align*}
|\mathrm{M}_{4,a}(\epsilon,\gamma)|=\begin{cases}
\frac{(p-1)^2p^{m}-(p-1)G_m\big(p^2-I_{2}(a)\big)}{2p^2},&\text{~if~}a\in\mathbb{F}_{p^m}\backslash\mathbb{F}_{p^2};\\	
\frac{(p-1)p^{m+1}-(p-1)G_m\big(p^2-I_{2}(a)\big)}{2p^2}, &\text{~if~}a\in\mathbb{F}_{p^2}\backslash\mathbb{F}_{p}.
\end{cases}
\end{align*}

		$(7)$  If  $\eta_m(b)\eta_{1}\big(\mathrm{Tr}(a^2b^{-1})\big)\eta_{1}\Big(-\frac{\mathrm{Tr}(a^2b^{-1})\mathrm{Tr}(b^{-1})-\mathrm{Tr}(ab^{-1})^2}{\mathrm{Tr}(a^2b^{-1})}\Big)=1$, then
	\begin{align*}
	|N_{a}(b,0)|=
	p^{m-3}-\frac{G_m}{p^2},
	\end{align*} and  the frequency is \begin{align*}
	|\mathrm{M}_{4,a}(\epsilon,\gamma)|=\begin{cases}
	\frac{(p-1)^2p^{m}+(p-1)G_m\big(p^2-I_{2}(a)\big)}{2p^2},&\text{~if~}a\in\mathbb{F}_{p^m}\backslash\mathbb{F}_{p^2};\\	
	\frac{(p-1)p^{m+1}+(p-1)G_m\big(p^2-I_{2}(a)\big)}{2p^2}, &\text{~if~}a\in\mathbb{F}_{p^2}\backslash\mathbb{F}_{p}.
	\end{cases}
	\end{align*}
		
		Note that the weight for the codeword $\mathbf{c}(b)\in\mathcal{C}_{D_a}$ is $p^{m-2}-N_{a}(b,0)$, by the above discussions and Lemma \ref{l22}, we can get the weight distribution for $\mathcal{C}_{D_a}$ directly. 
		
		$\hfill\Box$
\section{Applications}
 It is well-known that few-weight linear codes have better applications in secret sharing schemes \cite{JY2006,CC2005},  in particular, projective three-weight codes are very interesting since they are closely related to $s$-sum sets $(s\in\mathbb{Z}^{+},~s>1)$ \cite{SS,SK}. Here, we present the following two applications.
 \subsection{Applications for secret sharing schemes}
The secret sharing schemes is introduced by Blakley \cite{BG} and Shamir \cite{KK} in 1979. Based on linear codes, many secret sharing  schemes are constructed \cite{MJ,JY2006, MJ93, KD2015}. Especially, for those linear codes with any nonzero codewords minimal, their dual codes can be used to construct secret sharing schemes with nice access structures \cite{KD2015}. 

For a linear code $\mathcal{C}$ with length $n$, the support of a nonzero codeword $\mathbf{c}=(c_1,\ldots,c_n)\in\mathcal{C}$ is denoted by 
\begin{align*}
\mathrm{supp}(\mathbf{c})=\{i~|~c_i\neq 0,i=1,\ldots,n\}.
\end{align*} For $\mathbf{c}_1,\mathbf{c}_2\in\mathcal{C}$, when $\mathrm{supp}(\mathbf{c}_2)\subseteq\mathrm{supp}(\mathbf{c}_1)$, we say that $\mathbf{c}_1$ covers $\mathbf{c}_2$ . 
A nonzero codeword $\mathbf{c}\in\mathcal{C}$ is  minimal if it covers only the codeword $\lambda\mathbf{c}$ $(\lambda\in\mathbb{F}_p^{*})$, but no other codewords in $\mathcal{C}$. 

The following lemma is a criteria for the minimal linear code.
\begin{lemma}[Ashikhmin-Barg Lemma \cite{AA}]\label{51}
	Let $w_{min}$ and $w_{max}$ be the minimal and maximal nonzero weights of the linear code $\mathcal{C}$ over $\mathbb{F}_p$, respectively. If
	\begin{align*}
	\frac{w_{min}}{w_{max}}>\frac{p-1}{p},
	\end{align*}
\end{lemma}
then  $\mathcal{C}$ is minimal.\\

For $m\ge 6$, by Theorems $\ref{t1}$-$\ref{t3}$, we have the following three cases.  

$(1)$ If $m$ is odd and $a\in\mathbb{F}_{p^{m}}\backslash\mathbb{F}_{p}$, then
\begin{align*}
\begin{aligned}
\frac{w_{min}}{w_{max}}&=\frac{ p^{m-2}-p^{m-3}-p^{\frac{m-3}{2}}}{p^{m-2}-p^{m-3}+p^{\frac{m-3}{2}}}= 1-\frac{2}{p^{\frac{m-1}{2}}-p^{\frac{m-3}{2}}+p}	> \frac{p-1}{p}.
\end{aligned}
\end{align*}

$(2)$ If $m$ is even and $a\in\mathbb{F}_{p^{2}}\backslash\mathbb{F}_{p}$, then
\begin{align*}
\begin{aligned}
\frac{w_{min}}{w_{max}}&=\frac{ p^{m-2}-p^{m-3}-(p-1)p^{\frac{m-4}{2}}}{p^{m-2}-p^{m-3}+(p-1)p^{\frac{m-4}{2}}}= 1-\frac{2(p-1)}{p^{\frac{m}{2}}-p^{\frac{m-2}{2}}+(p-1)}	> \frac{p-1}{p}.
\end{aligned}
\end{align*}	 

$(3)$ If $m$ is even and $a\in\mathbb{F}_{p^{m}}\backslash\mathbb{F}_{p^2}$, then
\begin{align*}
\begin{aligned}
\frac{w_{min}}{w_{max}}&=\frac{ p^{m-2}-p^{m-3}-p^{\frac{m-2}{2}}}{p^{m-2}-p^{m-3}+p^{\frac{m-2}{2}}}= 1-\frac{2}{p^{\frac{m-2}{2}}-p^{\frac{m-4}{2}}+p}	> \frac{p-1}{p}.
\end{aligned}
\end{align*}

Hence, for $m\ge 6$, by Lemma \ref{51}, $\mathcal{C}_{D_a}$ is  minimal. Therefore, its dual code can be employed to construct secret sharing schemes with interesting access structures.

\subsection{Applications for $s$-sum sets}

In 1984, Courteau and Wolfmann introducted the notion of the triple-sum set\cite{SB}, a natural generalization of partial difference sets, which enter the study of two-weight codes \cite{RC1986,SM}. Recently, Shi and Sol$\acute{e}$ generalized the nonation of the triple-sum set to be $s$-sum set $(s>1)$, and then they gave the connection among $s$-sum sets, $s$-strongly walk-regular graphs and three-weight codes \cite{SS,SK}.

Some notations and results for the $s$-sum set are given as follows \cite{SS}.

For given integers $k$ and $s$,	a nonempty set $\Omega\subsetneq\mathbb{F}_{q}^{k}$ is an $s$-sum set  if it is stable by scalar multiplication, and for any nonzero $h\in\mathbb{F}_q^{k}$, there are constants $\sigma_0$ and $\sigma_1$ such that $h$ can be written as 
\begin{align*}
	h=\sum_{i=1}^{s}x_i~~~\big(x_i\in\Omega~\text{for}~i=1,\ldots,s\big)
\end{align*}with $\sigma_0$ times of $h\in\Omega$ and  $\sigma_1$ times of $h\in\mathbb{F}_q^{k}\backslash\Omega$.

We denote $\mathcal{C}(\Omega)$ by the projective code of length $n=\frac{|\Omega|}{q-1}$ with the parity check matrix $H$, where $H$ is the $k\times n$ matrix with columns  projectively non-equivalent  elements of $\Omega\backslash\{0\}$.

In the following, the connection between $\Omega$ and $\mathcal{C}(\Omega)$ is given.

\begin{lemma}[Theorem 6 in \cite{SS}]\label{lsss}
	Assume that $\mathcal{C}(\Omega)^{\perp}$  is of length $n$ and has three nonzero weights $w_1<w_2=\frac{n(q-1)}{q}<w_3$, with $w_1+w_3=\frac{2n(q-1)}{q}$. Then $\Omega$ is an $s$-set sum for any odd $s>1$. 
\end{lemma}

Now, we can construct an $s$-sum set for any odd $s>1$ from $\mathcal{C}_{D_a}$ in Theorem \ref{t1}.

For $n=p^{m-2}$ and $D_a=\{d_1,\ldots,d_{n}\}$, let $v_1,\ldots,v_m$ be a basis of $\mathbb{F}_{p^{m}}$ over $\mathbb{F}_p$, then $\mathcal{C}_{D_a}$ has an generator matrix
\begin{align*}
	G=\big(\mathbf{g}_1,\ldots,\mathbf{g}_n\big),
\end{align*} 
where 
\begin{align*}
	\mathbf{g}_i=\big(\mathrm{Tr}(v_1d_i^2),\ldots,\mathrm{Tr}(v_md_i^2)\big)^{\mathrm{T}}\quad(i=1,\ldots,n).
\end{align*}

Let \begin{align*}
	\Omega=\{\lambda\mathbf{g}_i~|~\lambda\in\mathbb{F}_{p}^*,~i=1,\ldots,n\}\subsetneq \mathbb{F}_p^{m}.
\end{align*} 
Now for any odd $m$, we prove that $\Omega$ is an $s$-set sum for any odd $s>1$.

By Theorem \ref{c5}, we know that $\mathcal{C}_{D_a}$ is projective, thus $\mathbf{g}_i\neq\lambda \mathbf{g}_j\!$~$(i\neq j,\lambda\in\mathbb{F}_p^*)$, and then, the weight distribution for $\mathcal{C}(\Omega)$ is the same as that for $\mathcal{C}_{D_a}^{\perp}$.
Now by Theorem $\ref{t1}$, we know that $\mathcal{C}(\Omega)^{\perp}$ is a  three-weight code with  three nonzero weights $w_1<w_2=\frac{n(q-1)}{q}<w_3$ and $w_1+w_3=\frac{2n(q-1)}{q}$. Now from Lemma \ref{lsss},  $\Omega$ is an $s$-set sum for any odd $s>1$. 

	 \section{Conclusions}
 In this paper, several classes of few-weight linear codes are constructed, and then their complete weight enumertors are determined by employing Gauss sums and quadratic character sums. Furthermore, some of these codes can be suitable for applications in secret sharing schemes and  $s$-sum sets for any odd $s>1$.

	 }
       \end{document}